\documentclass[times,review,preprint,authoryear]{elsarticle}

\usepackage{hyperref}
\usepackage{xr-hyper}

\usepackage{amssymb}
\usepackage{latexsym}
\usepackage{amsmath}

\usepackage{url}
\usepackage{xcolor}

\definecolor{newcolor}{rgb}{.8,.349,.1}

\usepackage{graphicx}
\usepackage{paralist}
\usepackage{siunitx}
\DeclareSIUnit\pixel{px}
\usepackage{booktabs}
\usepackage{appendix}
\usepackage{printlen}
\uselengthunit{mm}

\usepackage{color, colortbl}
\definecolor{Gray}{gray}{0.9}
\definecolor{LightCyan}{rgb}{0.88,1,1}

\usepackage{rotating}

\usepackage{subcaption}
\usepackage{afterpage}
\usepackage{everypage}

\usepackage[flushleft]{threeparttable}
\usepackage{tikz}
\usetikzlibrary{shapes.geometric}
\usetikzlibrary{calc}
\usetikzlibrary{positioning}
\usetikzlibrary{arrows.meta}
\usetikzlibrary{decorations.pathreplacing}
\usetikzlibrary{fit}
\usetikzlibrary{quotes}
\usetikzlibrary{fadings}
\usetikzlibrary{backgrounds}

\newcommand*{\BraceAmplitude}{0.16em}%
\newcommand*{\VerticalOffset}{0.8ex}%
\newcommand*{\HorizontalOffset}{0.0em}%
\NewDocumentCommand{\InsertRightBrace}{%
    O{} %
    O{\HorizontalOffset,\VerticalOffset} %
    m   %
    m   %
    m   %
}{%
    \begin{tikzpicture}[overlay,remember picture]
        \coordinate (Brace Top)    at ($(#3.north) + (#2)$);
        \coordinate (Brace Bottom) at ($(#4.south) + (#2)$);
    \draw [decoration={brace, amplitude=\BraceAmplitude}, decorate, thick, draw=blue, #1]
            (Brace Top) -- (Brace Bottom)
            node [pos=0.5, anchor=west, align=left, text width=1.25cm, color=black, xshift=\BraceAmplitude] {#5};
    \end{tikzpicture}%
}%

\usepackage{multirow}
\newcommand\betweenmodels{1ex}

\usepackage{multibib}
\usepackage{multicol}

\newcolumntype{P}{>{\scriptsize}p{0.35cm}}

\begin{document}

\begin{frontmatter}

\title{Multi-scale fully convolutional neural networks for histopathology image segmentation: from nuclear aberrations to the global tissue architecture}

\author[1,3,4]{R\"udiger Schmitz\corref{cor1}}
\author[3,4]{Frederic Madesta}
\author[3,4]{Maximilian Nielsen}
\author[5]{Jenny Krause}
\author[6]{Stefan Steurer}
\author[3,4]{Ren{\'e} Werner\corref{cor2}}
\author[1]{Thomas R\"osch\corref{cor2}}

\cortext[cor1]{Corresponding author:
  \texttt{r.schmitz@uke.de}.}
\cortext[cor2]{Equal contribution.}

\address[1]{Department for Interdisciplinary Endoscopy,
}
\address[3]{Center for Biomedical Artificial Intelligence (bAIome),
}
\address[4]{Department of Computational Neuroscience,
}
\address[5]{I. Department of Internal Medicine, %
}
\address[6]{Department of Pathology,
University Medical Center Hamburg-Eppendorf, Hamburg, Germany
}

\begin{abstract}
Histopathologic diagnosis relies on simultaneous integration of information from a broad range of scales, ranging from nuclear aberrations ($\approx \mathcal{O}(\SI{0.1}{\micro\metre})$) through cellular structures ($\approx \mathcal{O}(\SI{10}{\micro\metre})$) to the global tissue architecture ($\gtrapprox \mathcal{O}(\SI{1}{\milli\metre})$).

 To explicitly mimic how human pathologists combine multi-scale information, we introduce a family of multi-encoder fully-convolutional neural networks with deep fusion.
 We present a simple block for merging model paths with differing spatial scales in a spatial relationship-preserving fashion, which can readily be included in standard encoder-decoder networks.  %
 Additionally, a context classification gate block is proposed as an alternative for the incorporation of global context.

Our experiments were performed on three publicly available whole-slide images of recent challenges (PAIP 2019: hepatocellular carcinoma segmentation; BACH 2020: breast cancer segmentation; CAMELYON 2016: metastasis detection in lymph nodes). The multi-scale architectures consistently outperformed the baseline single-scale U-Nets by a large margin.
They benefit from local as well as global context and particularly a combination of both.
If feature maps from different scales are fused, doing so in a manner preserving spatial relationships was found to be beneficial.
Deep guidance by a context classification loss appeared to improve model training at low computational costs.
All multi-scale models had a reduced GPU memory footprint compared to ensembles of individual U-Nets trained on different image scales.
Additional path fusions were shown to be possible at low computational cost, opening up possibilities for further, systematic and task-specific architecture optimization.

The findings demonstrate the potential of the presented family of human-inspired, end-to-end trainable, multi-scale multi-encoder fully-convolutional neural networks to improve deep histopathologic diagnosis by extensive integration of largely different spatial scales.

\end{abstract}


\cortext[cor3]{\textbf{Abbreviations}: \textit{Arch.} -- architecture, \textit{CI} -- confidence interval, \textit{Clss.} -- classification, \textit{Ens.} -- enseble, \textit{FCN} -- fully-convolutional neural net, \textit{HCC} -- hepatocellular carcinoma, \textit{Mem.} -- GPU memory footprint (in GB per patch), \textit{ms} -- multi-scale, %
	\textit{msM} -- multi-scale merge-block,
	\textit{pms.} -- parameters,
	\textit{SVM} -- support vector machine,
	\textit{WSI} -- whole-slide image.}

\end{frontmatter}

\section{Introduction}
\label{sec:introduction}

\subsection{Clinical relevance and motivation}

\textit{If the rumor is tumor, the issue is tissue.} Histopathology is the gold standard and backbone of cancer diagnosis, providing important information for various stages of the treatment process \citep{uicc_tnm_8thEd}. For instance, a fine-grained grading and staging of dysplasia and malignancy in precursor and cancer lesions, respectively, underlies individualized treatment planning in many tumor entities. Moreover, in curative surgery, the assessment of whether the resection specimen margins are free of tumour cells is of vital importance and a core task of clinical pathology.

Human pathologists meet these challenges with the help of elaborated diagnostic criteria and grading systems for all kinds of cancer and cancer precursors. Even though their specific details vary for different kinds of cancer, many rely on a combination of features such as
\begin{itemize}
\item Nuclear inner density, i.e. color, \citep{Zink2004} and
\item Deformed and varying nuclear shapes or global alterations of the nuclei \citep{Zink2004}.
\item Increased nucleus to stroma ratio and
\item Loss of nuclear polarity (e.g. nucleus not anymore at the bottom), as observed in many glandular tumours.
\item Deformed cellular shapes and heterogeneous cell sizes,
\item Loss of organ- and function-defining positions on small scales (e.g., neighboring cells not in a single layer, but some stacked over each other) and larger scales (e.g., atypical or deformed glandular shapes),
\item Invasion (i.e., disrespecting global tissue order and borders between different layers).
\end{itemize}

As can be seen from this (not-exhaustive) list, diagnosis and grading of malignancy inherently involve a range of different scales. These scales may span a factor of more than a thousand-fold, ranging from sub-nuclear features (which lie on a spatial scale of $\approx \mathcal{O}(\SI{0.1}{\micro\metre})$) via nuclear, cellular ($\approx \mathcal{O}(\SI{10}{\micro\metre})$), inter-cellular ($\approx \mathcal{O}(\SI{100}{\micro\metre})$) to glandular and other higher organisational features ($\gtrapprox \mathcal{O}(\SI{1}{\milli\metre})$).
The importance of the integration of information from different scales is reflected in how human pathologists approach these tasks: Regions of interest are repeatedly viewed at several different magnifications by turning the objective revolver of the microscopy back and forth.

In this work, we aim to develop a family of deep learning models that architecturally mimic this behaviour. %

\subsection{Related works}

With their success in various computer vision tasks, deep learning methods have opened up a myriad of perspectives for computer vision and computer-aided diagnosis (CADx) in histopathology \citep{Litjens2017}. Image segmentation is a standard task in computer vision and machine learning and has a direct clinical use in the field of pathology, be it the analysis of the margin status (i.e., distance of tumor cells to resection margin), area-dependent grading systems (with the Gleason score in prostate cancer as a prominent example \citep{Gleason,karimi2020DLGleason,nir_automatic_2018}), or specific research applications, such as the analysis of 3d-tumor morphology from a multiplicity of tissue sections \citep{Schmitz2018, Segovia2019}.

In medical image segmentation, standard computer vision models, including fully convolutional neural networks (FCNs, \citealt{Long2015}) and, most prominently, U-Net-based architectures \citep{Ronneberger2015} have successfully been applied to various scenarios and imaging modalities \citep{Litjens2017, Isensee2018}, including computational histopathology \citep{Bulten2019, Liu, Campanella2019}.

In addition, more specialized network architectures \citep{Bejnordi2017, li_multi-scale_2018, Vu2019} and training techniques \citep{Campanella2019, Wang2019} have been proposed to address the challenges of computational histopathology.
Some of these works touch upon the question on how additional context can be provided to the network, but are mainly confined to local, similar-scale context \citep{Bejnordi2017, li_multi-scale_2018} and/or sliding-window convolutional neural network (CNN) techniques \citep{Bejnordi2017, wetteland_multiscale_nodate} or classification tasks.
Early on, \citet{nir_automatic_2018} described the potential of multi-scale features from more separate scales in classical, "hand-crafted" feature-driven machine learning.
By integrating the features from different scales by use of a support vector machine (SVM), they paved the way for many works to follow. Recently, the same research group advocated the use of individually trained CNNs as feature extractors whose outputs were, in an ensembling-like fashion, combined by a logistic regression model into a final Gleason grade classification in prostate cancer \citep{karimi2020DLGleason}. Similarly,  \citet{wetteland_multiscale_nodate} suggested to train distinct CNNs as feature extractors and merge their information in a classification network in replacement of the original fully-connected layers, which can again be viewed as an ensembling approach. For application to breast cancer and its differential diagnoses, \citet{ning_multiscale_2019} proposed a similar ensembling technique, but again based on classical, hand-crafted feature extractors and using an SVM for integration of multi-scale information.

In 3d imaging, multi-path end-to-end trainable models have incorporated similar-scale and local context to reduce the memory footprint and alleviate the problem of the otherwise extremely limited input size in memory-costly 3d-nets \citep{Kamnitsas2017}, with remarkable success in e.g. the sub-acute stroke lesion segmentation challenge ISLES 2015 \citep{maier_isles_2017}.
In histopathology, there also have been attempts toward end-to-end trainable multi-scale models  \citep{gu_multi-resolution_2018,li_multi-scale_2018}, but which have so far shown only minor benefits as compared to the aforementioned ensembling variants. This observation indicates that multi-scale deep learning-based segmentation of histopathology data is still in its infancy and its actual potential remains to be unveiled.

Introduction of additional image context has been a topic of interest also in the natural image domain, resulting in prominent techniques like dilated or atrous convolutions and atrous spatial pyramid pooling \citep{Chen2017DeepLab, Chen2017Atrous}. %
Further, \citet{Zhang2018} proposed an FCN architecture that explicitly predicts "scaling factors" for the possible classes from the bottle-neck layer, which are then used to multiply and, thus, highlight the respective feature maps at the final layer. The scaling factors can capture the overall image content and can be trained by use of an additional classification loss.
Similarly, \citet{Zhou2019} introduced a reinforcement-based strategy involving two sub-nets, one for encoding context and one for the actual segmentation task.
By the properties of the natural image domain, namely the limited image size as compared to histopathologic whole-slide images (WSIs), the scales of detailed and contextual features in these works are, however, much more similar than in histopathology.
Congruously, the primary aim of these approaches has been to "help clarify local confusion" \citep{Liu2015} or to make better use of what is fed into to the net anyways, rather than to add large and otherwise unavailable context.
For histopathology image segmentation as a specific task, however, we aim for the integration of otherwise unavailable information from much different scales into a single, end-to-end trainable model.

\subsection{Contributions}

There exist plenty of highly optimized, U-Net-derived architectures employing, for instance, elaborate skip connections \citep{badrinarayanan_segnet_2017}, dense connections \citep{li_denseunet_2018}, attention gating techniques \citep{oktay_attention_2018} and newer FCN architectures like DeepLab \citep{Chen2017DeepLab}.
Nevertheless, standard U-Nets have turned out to be robust work horses for many medical computer vision tasks and are hard to beat by internal modifications of the base architecture \citep{Isensee2018, isensee_breaking_the_spell_2019}.

However, histopathology diagnosis is, by the nature of the large whole-slide images and with closely interwoven features from very different scales, a very specific and challenging task.
Therefore, drawing on the U-Net as a standard base model, this work explores whether an architectural mimicry of how human experts approach this specific task can improve the performance of FCNs for histopathology image segmentation.

The main contributions of this paper are as follows:
We introduce a family of U-Net-based fully convolutional deep neural nets that are specifically designed for the extensive integration of largely different spatial scales.
First, we propose a simple building block that can fuse various encoders with different spatial scales in a manner that preserves relative spatial scales. As a light-weight alternative, we also propose the use of an independent context classification model for gating the segmentation model output.
Second, we integrate these building blocks
into different multi-scale FCNs and %
compare their segmentation performance to U-Net baseline architectures. To illustrate generalizability, the evaluation is based on three different publicly available image datasets provided by recent challenges.
Third, by a systematic, stepwise analysis, we identify relevant aspects of the proposed multi-scale FCN family, including the necessity of preserving spatial relationships between different encoders, the benefits from deep guidance by an additional classification loss and possible generalizations through multiple path fusions.
Based on these observations, we narrow down the possible multi-scale setups and comment on how to systematically adapt the presented multi-scale FCN family to specific deep learning tasks in histopathology.

To foster reproducibility and further research, our proposed models are publicly provided as open source\footnote{https://github.com/ipmi-icns-uke/multiscale/. Please do not hesitate to contact the corresponding author for help with implementation and usage.\label{githubour}} to the community.

\section{Model architectures}

\subsection{Baseline architectures}

\subsubsection{U-Net architecture}

Beyond the still popular sliding-window CNN-based techniques, U-Net-based FCN architectures form the de facto standard in the medical image domain \citep{Isensee2018}, including histopathology \citep{Litjens2017}.
Beating the standard U-Net by internal modifications is evidently hard \citep{isensee_breaking_the_spell_2019, Isensee2018} and beyond the scope of this work.
Rather, as outlined in the introduction, this study examined whether by designing a model of standard components but with its larger architecture mimicking human expert diagnostic procedures, further improvement can be made.

Therefore, we chose a non-modified ResNet18-based U-Net \citep{He2015} as a common, standard U-Net variant to form the baseline for this study\footnote{
An implementation of this architecture can be found at %
https://github.com/usuyama/pytorch-unet. Accessed: 2019-09-19.}.
For a detailed description of the baseline model, the reader is referred to section \ref{sec:supp_baseline_results} in the Supplementary Materials.
In brief, the ResNet18 forms the encoder of the otherwise standard U-Net architecture (cf. figure \ref{fig:supp_baseline}), where the encoding ResNet18 has been pre-trained on the ImageNet dataset \citep{ImageNet}.
For our study, the baseline model was trained at full-resolution patches of $512 \times 512$ pixels of the WSI images.

\subsubsection{Multi-scale ensembles as an upper bound for state-of-the-art performance}

We additionally compared our proposed models to ensembling techniques to see if we can reach or even excel their performance with a single, and computationally less costly, model. Inspired by the approach of \citet{karimi2020DLGleason} to Gleason grading, individual U-Nets were trained on the different image scales. The predicted probabilities were then re-sampled to the target resolution and merged by use of different ensembling techniques, %
namely hard majority votes, average ensembles (soft majority voting), and logistic regression ensembles. For each ensemble, the best individual model per scale was selected. The logistic regression model was trained and evaluated on the same train and test sets as the individual models. By this procedure, %
we aimed to define a thorough and systematic upper bound for state-of-the-art multi-scale ensembling performance.

\subsection{The msY model family: Multi-scale multi-encoder architectures}
\label{sec:multi_scale_models}

\begin{figure*}[htb]
    \centering
    \includegraphics[width=\textwidth]{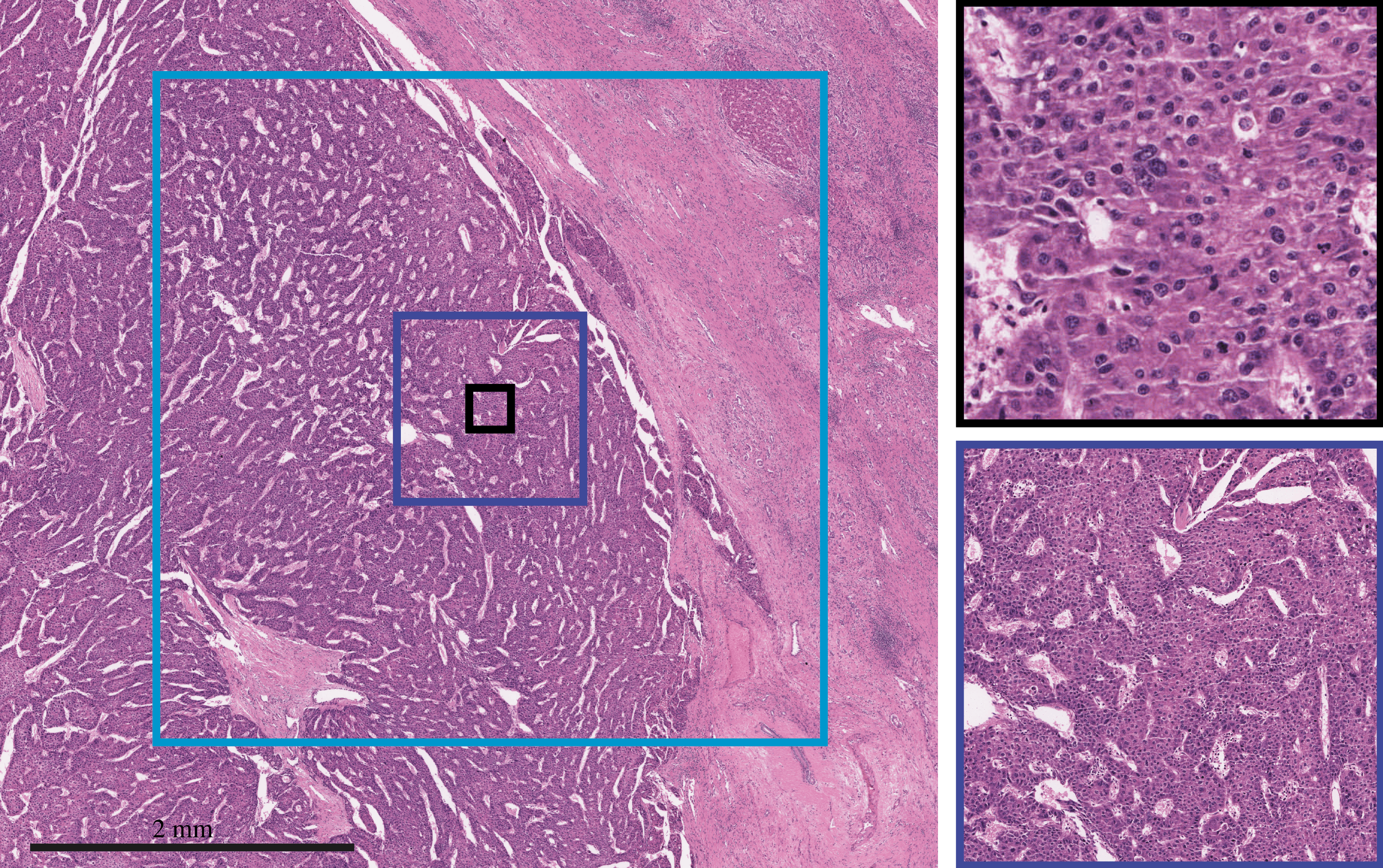}
    \caption{Input patches at different spatial scales, shown in an illustrative region of the same whole-slide image as depicted in figure \ref{fig:dataset_example}.
    The innermost, black rectangle corresponds to a $512 \times 512$ pixel patch of the scale 1, which we refer to as the ``detail patch'' (zoomed view in the top right inset).
    The next, dark blue rectangle corresponds to a $572 \times 572$ pixel patch of the scale 4. It contains information on how the cells are organised in strands and "trabeculae" -- or whether the cells violate these patterns. These features are hard or impossible to deduce using the innermost patch alone. In this sense, the dark blue patch adds "architectural" information. We refer to it as "local context" or the "local context patch". A zoomed view of it is shown in the bottom right inset.
    The outermost, light blue rectangle, which we call a "global context patch", contains information on the large-scale organization of the tissue, such as the pseudocapsule, which is typical for hepatocellular carcinoma.
    Whilst a standard U-Net is provided with the information from the detail patch solely, a msY-Net architecture (section \ref{sec:msY-Net}) can integrate information from the detail patch plus either the local or the global context patch. The msY$^2$- and msY$\textit{I}$-Net architectures are two options for integration of all three scales.
    The scale bar is 2 mm.}
    \label{fig:patches}
\end{figure*}

To provide the network with context and architectural information (cf. figure \ref{fig:patches} and the considerations in section \ref{sec:introduction}), we constructed a family of multi-scale multi-encoder networks building upon the baseline Res-U-Net architecture.

In the following, we first introduce the underlying blocks for integration of multi-scale context, namely the \textit{multi-scale merge block} and the \textit{context classification gate}.
Afterwards, we describe the different variants of our setup that are examined in this study.

Technically, it is worth noting that common whole-slide image (WSI) formats use so-called pyramid representations, which contain the original image in multiple, downsampled versions.
Therefore, multiple scales can directly be loaded from file, with no need for resampling and, hence, only a moderate overhead only.

\subsubsection{Multi-scale merge block: spatial relationship preserving path fusion in multi-scale multi-encoder models}
\label{subsec:merge_block}

\begin{figure}[htb!]
    \begin{subfigure}{\linewidth}
    \centering
    \includegraphics[width=\linewidth]{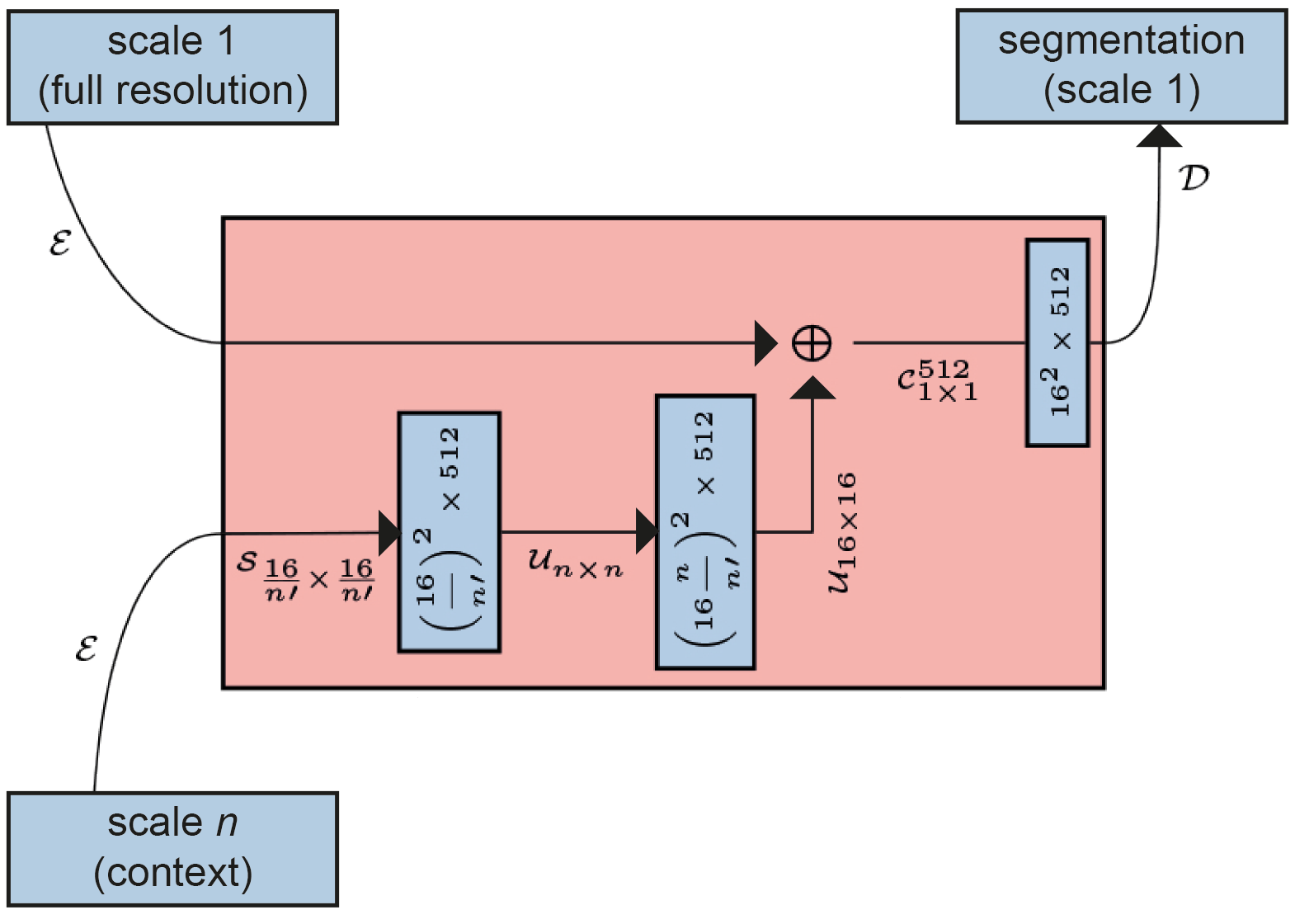}
    \caption{Multi-scale merge block}
    \vspace{0.5cm}
    \label{subfig:msM-Block}
    \end{subfigure}
    \begin{subfigure}{\linewidth}
    \includegraphics[width=\linewidth]{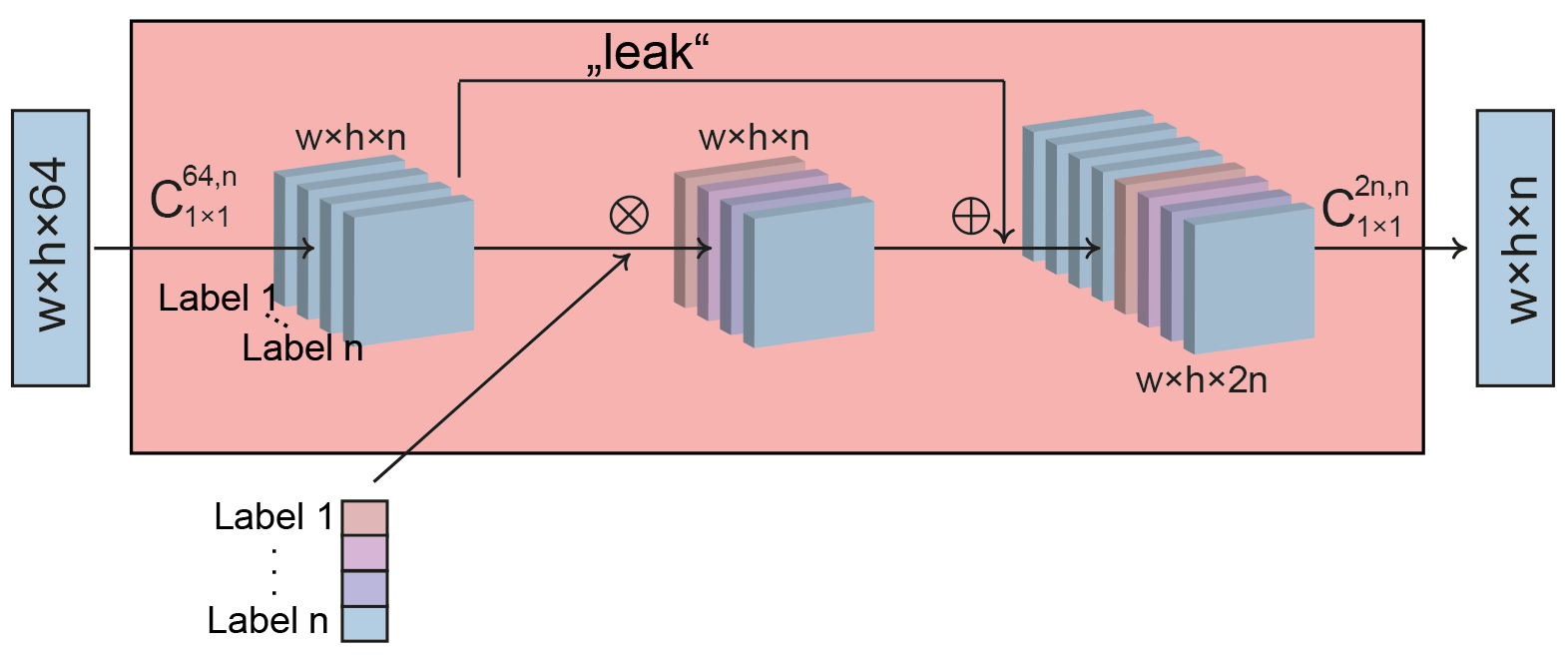}
    \caption{Context classification gate block}
    \label{subfig:ccG-Block}
    \end{subfigure}
    \caption{
    Schematic illustration of the multi-scale merge block (a) and the context classification gate block (b). $\mathcal{E}$, $\mathcal{D}$ are the encoder and decoder path of the network (cf. figure~\ref{subfig:YNetArchitecture}), respectively.
    The blue upward boxes are the feature maps with their respective sizes printed inside.
    $\oplus$ denotes concatenation and $\otimes$ channel-wise multiplication.
    The "leak" connection is a copy, followed by concatenation.
    $\mathcal{S}_{x \times y}$ stands for the central cropping of the size $x \times y$ in the spatial dimensions and $\mathcal{U}_{a\times b}$ the bilinear upsampling by factors of $a$ and $b$ in the dimensions 1 and 2.
    $w$ and $h$ denote the spatial width and height, $n$ the number of classes.
    $\mathcal{C}^{a,b}_{n_1,n_2}$ means $n_1 \times n_2$ convolution with $a$ input feature maps and $b$ output features maps (where $a$ may be omitted if the size of its input is explicitly given), followed by ReLU activation.
    }
    \label{fig:blocks}
\end{figure}

Figure \ref{subfig:msM-Block} sketches the functioning of the multi-scale merge block.
At the bottleneck level, the feature maps from both the main encoder and the side (context) encoder have sizes of $16 \times 16 \times 512$.
In order to spatially match the output of the full resolution encoder, a $n^\prime \times n^\prime$ center cropping ($\mathcal{S}_{\frac{16}{n^\prime}\times \frac{16}{n^\prime}}$) of the $n$-times down-scaled context path is performed, followed by $n\times n$ bilinear upsampling ($\mathcal{U}_{n\times n}$), where $n^\prime = 4$ if $n=4$ and $n^\prime =8$ if $n=16$. For the case $n^\prime \neq n$, another center cropping with $16 \times 16$ is conducted. Both, now spatially consistent paths are then merged by concatenation. Finally, the number of feature maps is reduced to the original number by a $1\times 1$ convolution. This operation is meant to learn which of the feature maps from the two paths are relevant and how they need to be combined.

For application to multiple context encoders, spatial alignment is ensured for any individual context encoder in the same manner as described above. The spatially-aligned feature maps from all encoders are then concatenated. Afterwards, the feature map size is reduced by a $1 \times 1$ convolution with $512 \cdot (m+1)$ input feature maps and $512$ output feature maps, $c^{512 \cdot (m+1), 512}_{1 \times 1}$, where $m$ denotes the number of side encoders.

\subsubsection{Context classification gate}
\label{subsec:context_gate}

Alternatively to the use of multiple encoders that are merged by multi-scale merge blocks, we examined the potential of gating the class-wise predictions from the segmentation network by a context classifier.
The context classifier tries to predict the content of the central detail patch from the low-resolution global context patch alone (cf. figure \ref{fig:patches}). As multiple, even mutually exclusive classes can be included in the detail patch, this states a multi-label classification problem.

The context classification gate is depicted in figure \ref{subfig:ccG-Block}.
The classification net outputs a probability value for each individual class (illustrated by the colored boxes in the bottom left corner). These are multiplied to the probability maps of the segmentation network in a channel-wise manner (similar to how excitations in squeeze-and-excite blocks are handled, cf. \citep{He2015}), thereby emphasizing probable classes and suppressing unlikely diagnoses.
To allow the segmentation network to either use or ignore this guidance, a "leak" is constructed by concatenation of the original, un-excited feature maps to the excited ones, followed by a $1 \times 1$ convolution that is to learn how to combine the excited and the leaked feature maps.

\subsubsection{msY-Net: Integrating context and tissue architecture}
\label{sec:msY-Net}

The msY-Net is provided with two patches of the scales 1 and 4 (which correspond to the inner two rectangles in figure \ref{fig:patches}) or 1 and 16 (inner and outer rectangle) as input. The full-resolution patch (scale 1) is fed into the standard U-Net architecture. The other is passed through a separate but analogous encoder architecture ("context encoder") built from another ResNet18. As the skip connections in the U-Net are for helping the decoder re-localise the high-level features and only the full-resolution patch is the one that needs to be segmented, the context encoder does \textit{not} have any skip connections to the decoder.

The two paths are merged at the bottleneck of the original U-Net by use of the multi-scale merge block (cf. section \ref{subsec:merge_block}). The resulting architecture is sketched in figure \ref{subfig:YNetArchitecture}.

In the following, we refer to a msY-Net that uses detail patches of the scale 1 and context patches of the scale $n$ as msY$_{(n)}$-Net.

\subsubsection{msY\textit{I}-Net and msY$^2$-Net: Integration of global and local context}
\label{subsec:msYIandmsY2}

In order to provide the model with large- and small-scale context information at the same time, we constructed two models that either use two context encoders or one context encoder plus one context classification gate.
For the former variant, we added two context encoders using a single multi-scale merge block. We refer to this model as msY$^2$-Net.

\begin{figure*}
    \begin{minipage}[t]{0.425\textwidth}
        \vspace{0pt}
        \begin{subfigure}{\linewidth}
        \centering
        \includegraphics[width=\linewidth]{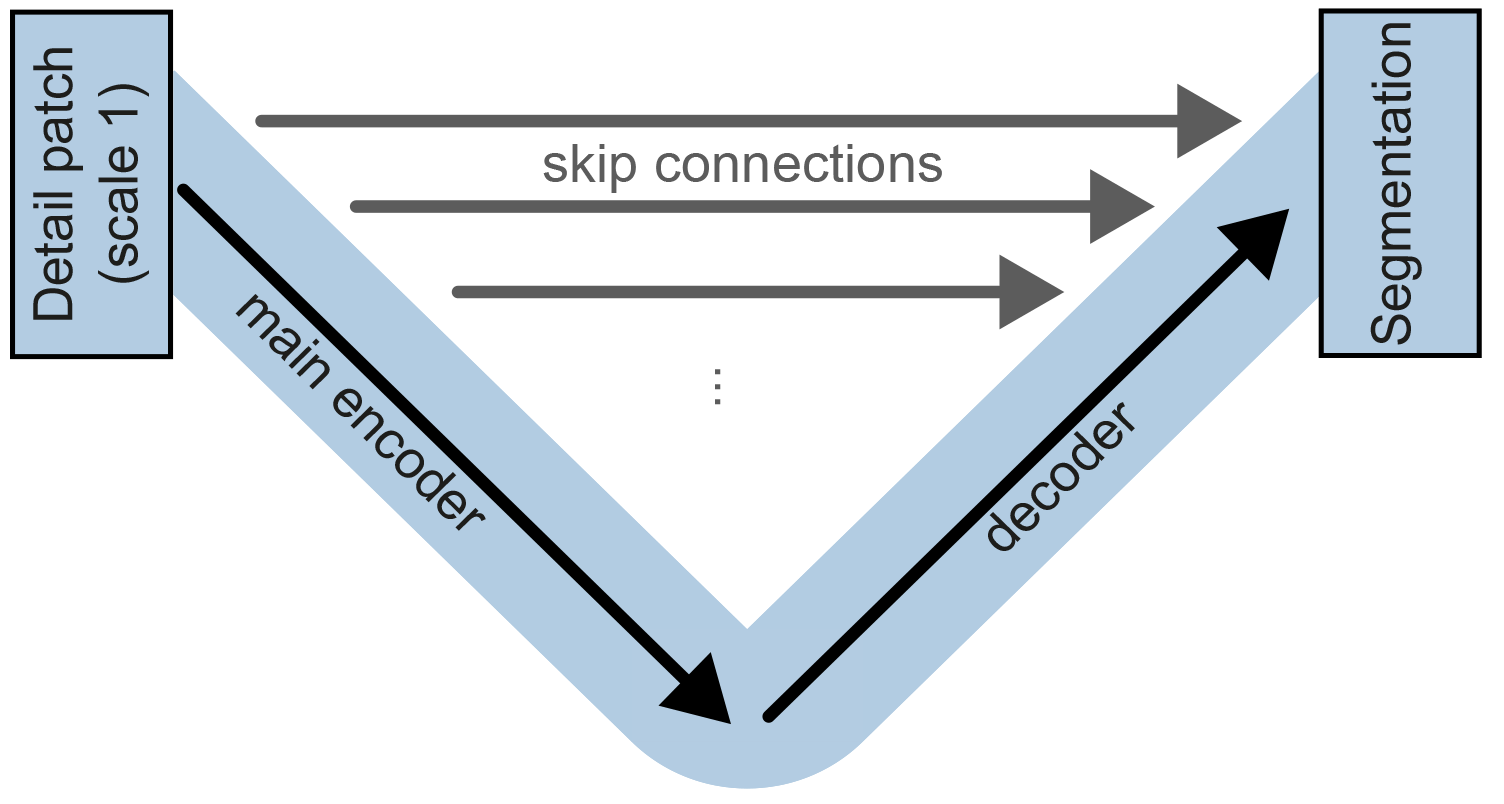}
        \caption{U-Net (baseline)}
        \label{subfig:UNetArchitecture}
        \end{subfigure}
        \begin{subfigure}{\linewidth}
        \centering
        \includegraphics[width=\linewidth]{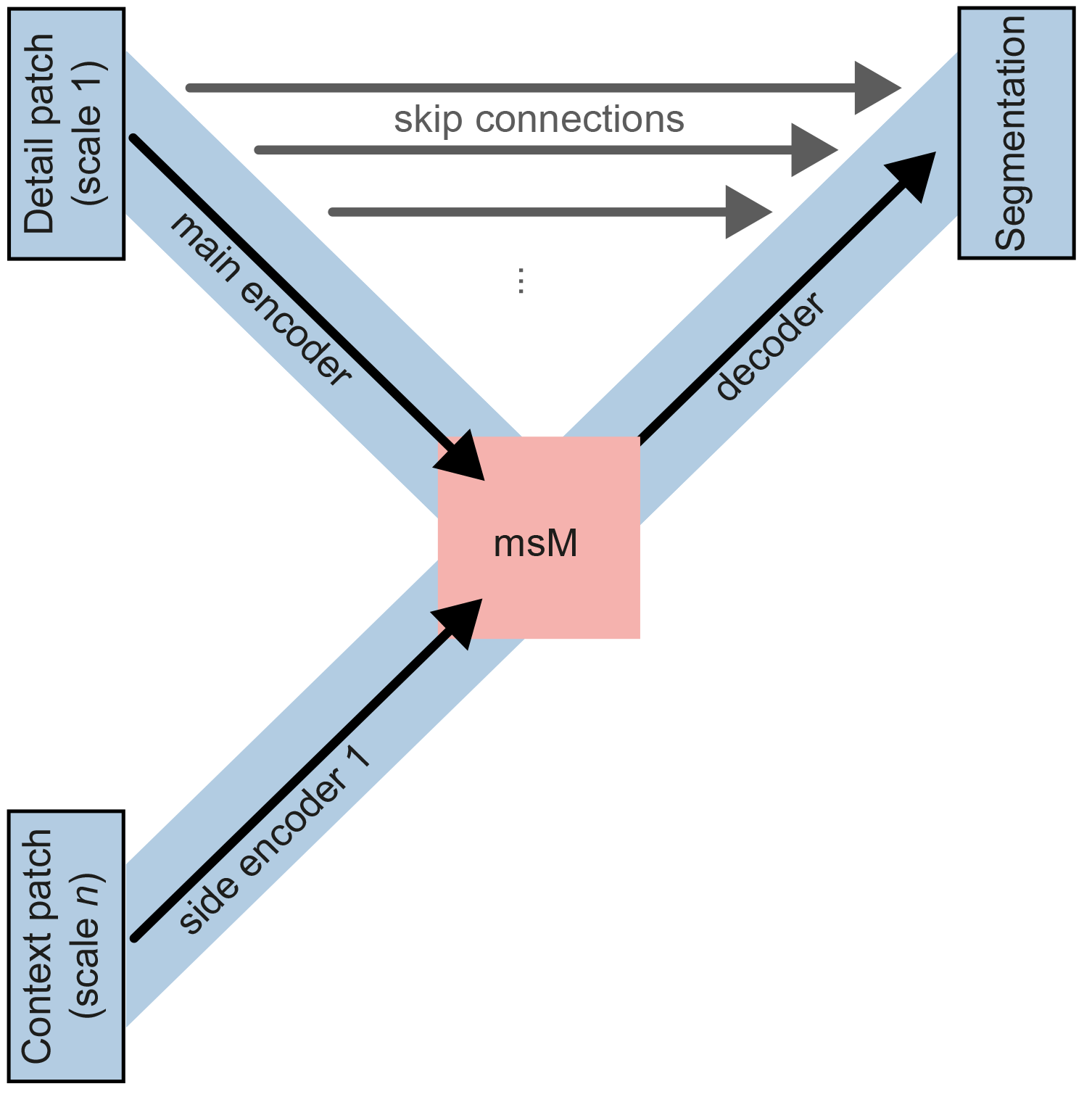}
        \caption{msY-Net}
        \label{subfig:YNetArchitecture}
        \end{subfigure}
        \begin{subfigure}{\linewidth}
        \centering
        \includegraphics[width=\linewidth]{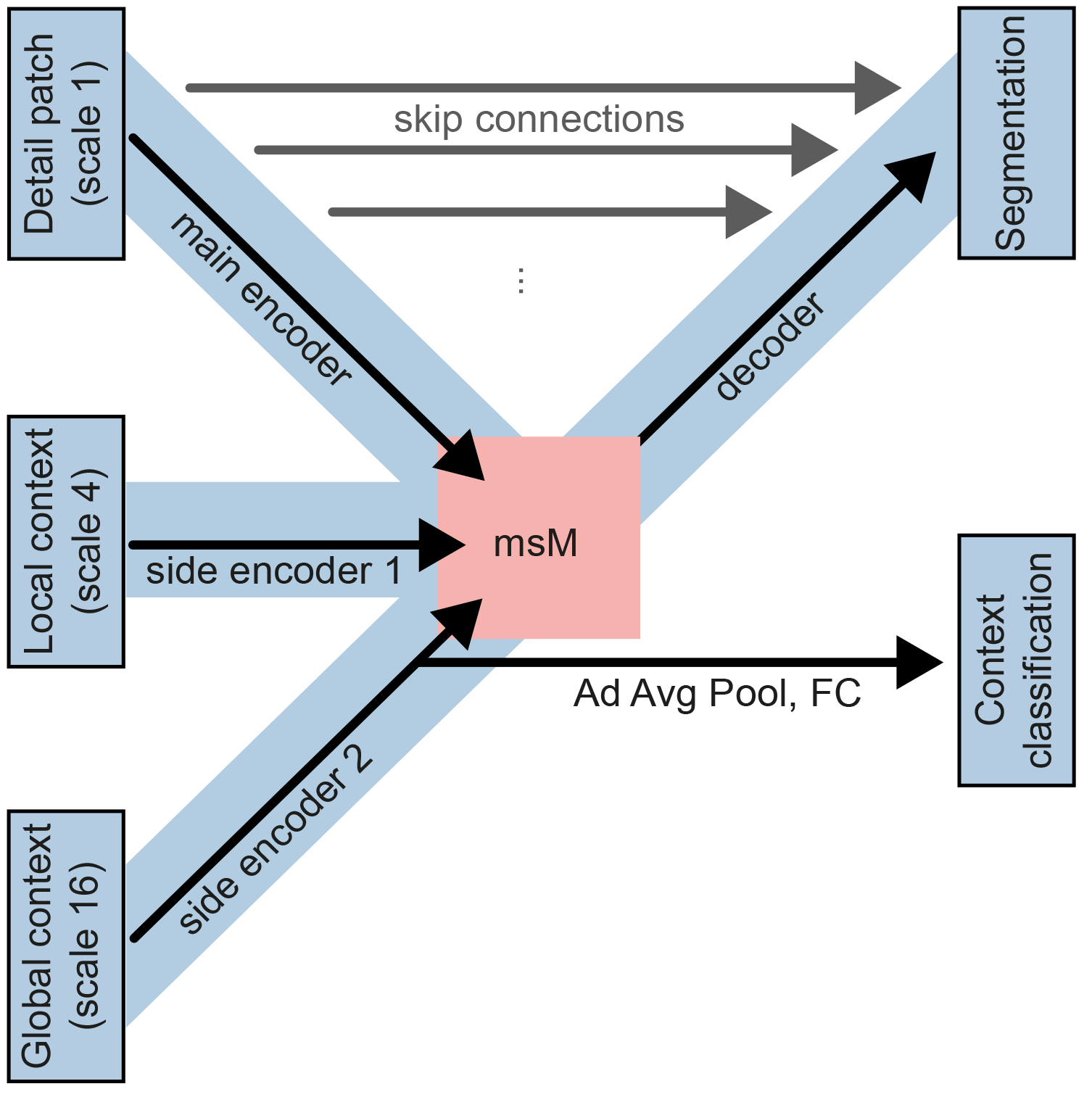}
        \caption{msY$^2$-Net}
        \label{subfig:Y2classNetArchitecture}
        \end{subfigure}
    \end{minipage}
    \hfill
    \begin{minipage}[t]{0.525\textwidth}
        \vspace{0pt}
        \begin{subfigure}{\linewidth}
        \centering
        \includegraphics[width=\linewidth]{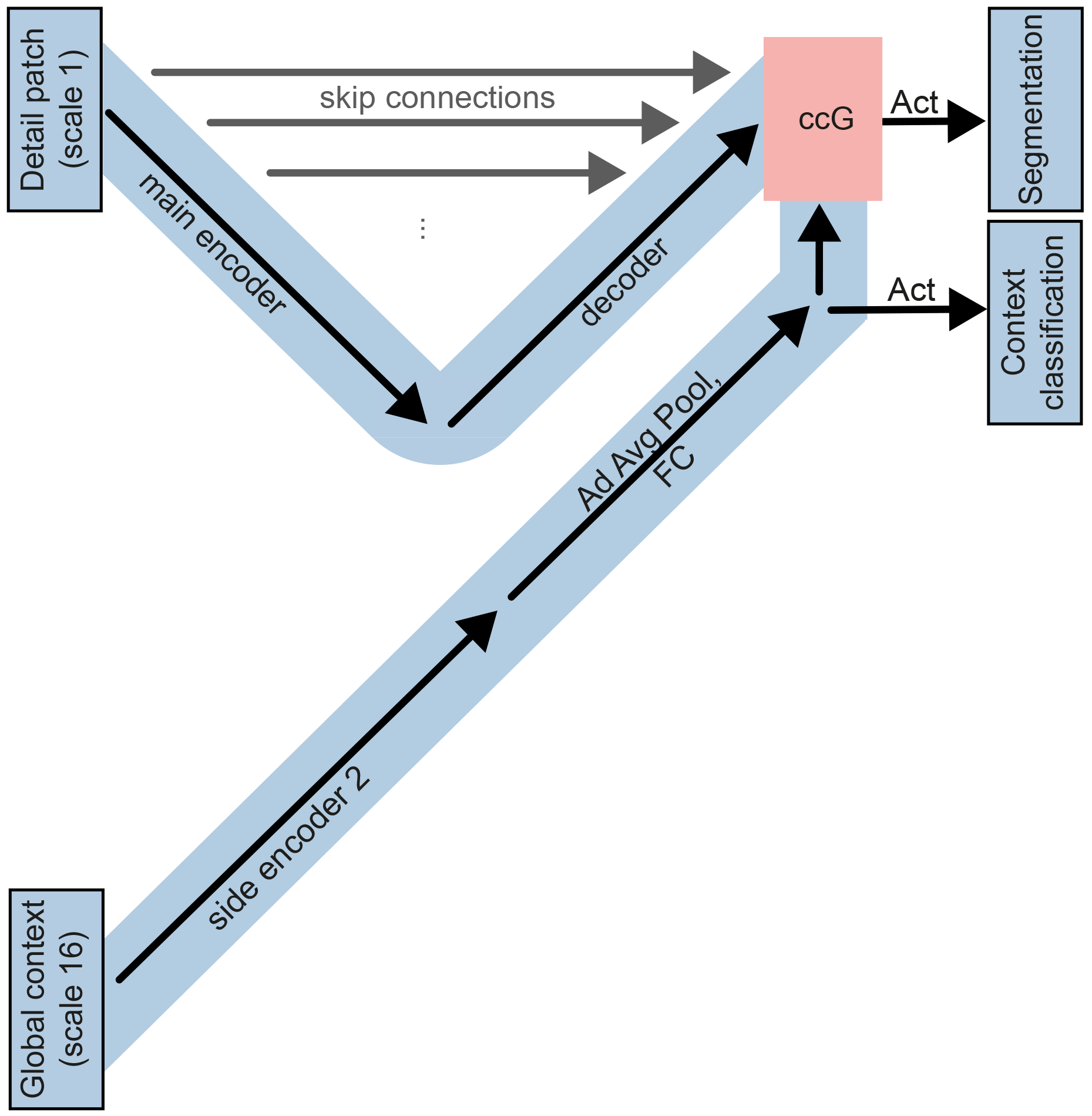}
        \caption{msU\textit{I}-Net}
        \label{subfig:UINetArchitecture}
        \end{subfigure}
        \begin{subfigure}{\linewidth}
        \centering
        \includegraphics[width=\linewidth]{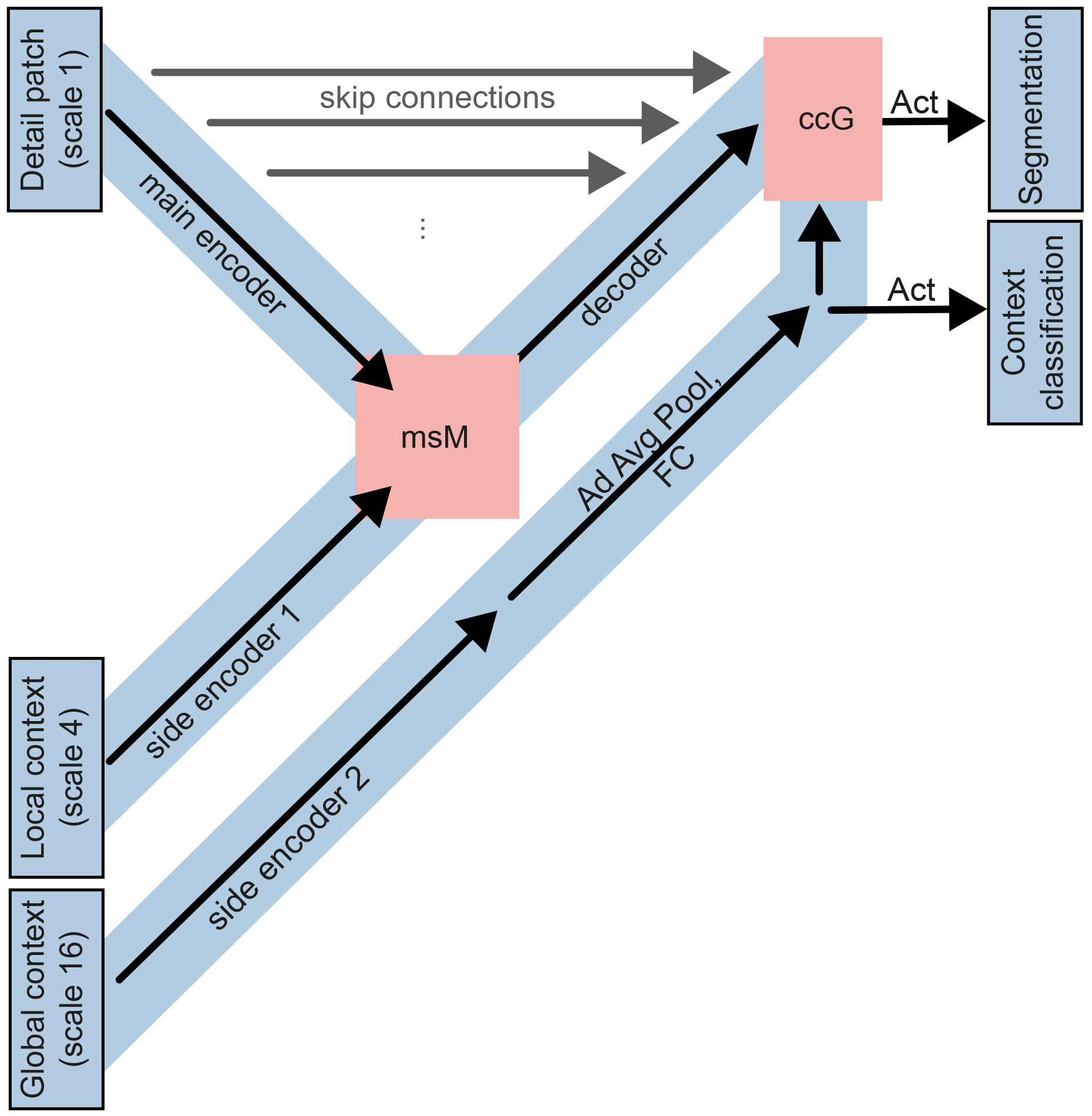}
        \caption{msY\textit{I}-Net}
        \label{subfig:YINetArchitecture}
        \end{subfigure}
        \caption{Schematic illustration of the main architectures studied in this paper. All main and side encoders use an ImageNet-pretrained ResNet-18. The skip connections and the decoder are the same throughout all of our models (see section \ref{sec:supp_baseline_details} for details). The multi-scale merge (msM) and context classification gate (ccG) blocks are described in sections \ref{subsec:merge_block} and \ref{subsec:context_gate} and sketched in figure \ref{fig:blocks}. Ad Avg Pool denotes adaptive average pooling, FC a fully-connected layer and Act the activation function.}
    \end{minipage}
    \label{fig:architectures}
\end{figure*}

For the latter variant, a context classification gate that uses
the large-context patch was added to an underlying msY-Net. This model is outlined in figure \ref{subfig:YINetArchitecture}. The \textit{I} in the name msY\textit{I}-Net refers to the large-context classification sub-net paralleling the underlying msY-Net \textit{without} any fusion at the bottleneck or before.
This network has two outputs: the segmentation of the full-resolution patch from its msY-Net part and the classification of the full-resolution patch content from the large-context encoder, its \textit{I}-part.
Finally, the final logits of the two paths are combined by a context classification gate that modifies the segmentation output by the classification of its context (cf. section \ref{subsec:context_gate}).

In the msY-Net and the msY$^2$-Net architectures, spatial correspondence between the full resolution encoder and the context encoder(s) is enforced in the multi-scale merge block (cf. figure \ref{fig:blocks}). It should be noted that for the large-context encoder in the msY\textit{I}-Net that ends in a classification gate instead of a multi-scale merge block, there is no such requirement. Therefore, this model can, in principle, be fed with large-context patches of arbitrary scales.

\subsubsection{Context classification loss}
\label{subsec:context_loss}

Additional loss functions can improve the training of specific parts of U-Net-based architectures and are used in various manners \citep{Kickingereder2019, Li2019}, including a classification loss on an additional output derived from the bottleneck feature maps \citep{Metha2018}.

Analogously, we computed a classification output from the feature maps of global context encoders (i.e., those with input scale 16)  and used it to compute an additional classification loss. The classification loss is computed with respect to the content of the detail patch.
As described in section \ref{sec:experiments}, we used a binary cross entropy (BCE) loss for the classification problem and added it to the segmentation loss.

By the additional classification loss, we wanted to ease gradient flow through the deeper layers of the context encoder and to explicitly force it to focus on the content of the detail patch.

\afterpage{\clearpage}
\AddThispageHook{\thispagestyle{empty}}

\subsubsection{Early and multiple fusions by multi-scale merge blocks}

Positioning the multi-scale merge block at the bottleneck level is not obligatory.
Technically, a multi-scale merge block can merge any two encoders at any level at which the one shall influence the other (the connection is directed, not mutual). Also, usage of multiple multi-scale merge blocks is possible.

Very early merge blocks will, however, not provide much context due to the small receptive fields of the center crop.
On the other hand, earlier and in particular multiple fusions may allow for an additional processing of combined features and might facilitate modelling of complex combined features.

It should be noted that these multiple merges are computationally inexpensive, as they only introduce additional learnable parameters through the $1 \times 1$ convolutions inside the merge blocks.

All models are implemented using PyTorch v1.2.0 \citep{Pytorch}.

\section{Materials \& methods}

\subsection{Datasets}

To examine generalizability of our findings, experiments were conducted on three different datasets for different entities of cancer, collected by different centers and scanned by different scanners.
To ensure reproduciblity, we employed the following three publicly available challenge datasets:

\subsubsection{PAIP 2019}
\label{sec:PAIP_description}
The PAIP (Pathology Artificial Intelligence Platform) 2019 challenge (part of the MICCAI 2019 Grand Challenge for Pathology) dataset comprises 50 de-identified  whole-slide histopathology images from 50 patients that underwent resection for hepatocellular carcinoma (HCC) in three Korean hospitals (SNUH, SNUBH, SMG-SNU BMC) from 2005 to June 2018 \citep{Paip2019Dataset}. The slides have been stained by hematoxylin and eosin and digitalized using an Aperio AT2 whole-slide scanner at $\times 20$ power and \SI[per-mode=symbol]{0.5021}{\micro\meter\per\pixel} resolution, resulting in image sizes between $35,855 \times 39,407$ and $64,768 \times 47,009$ pixels (1.399 to 3.044 gigapixels).

Regions of viable cancer cells as well as whole cancer regions (additionally including stroma cells and so forth) have been annotated manually. As described in \citep{Paip2019Dataset}, one pathologist with 11 years of experience in liver histopathology drew the initial annotations that were then reviewed by another expert pathologist. All cases of the given dataset include cancer regions. With respect to the Edmonson-Steiner grading system \citep{Edmondson1954}, their distribution is as follows: $N=7$ cases of grade 1, $N=23$ grade 2 tumors, and $N=20$ grade 3 samples.

All de-identified pathology images and annotations were prepared and provided by the Seoul National University Hospital under a grant from the Korea Health Technology R\&D Project through the Korea Health Industry Development Institute (KHIDI), funded by the Ministry of Health \& Welfare, Republic of Korea (grant HI18C0316).

\begin{figure}[h!]
    \centering
    \includegraphics[height=0.9\textheight]{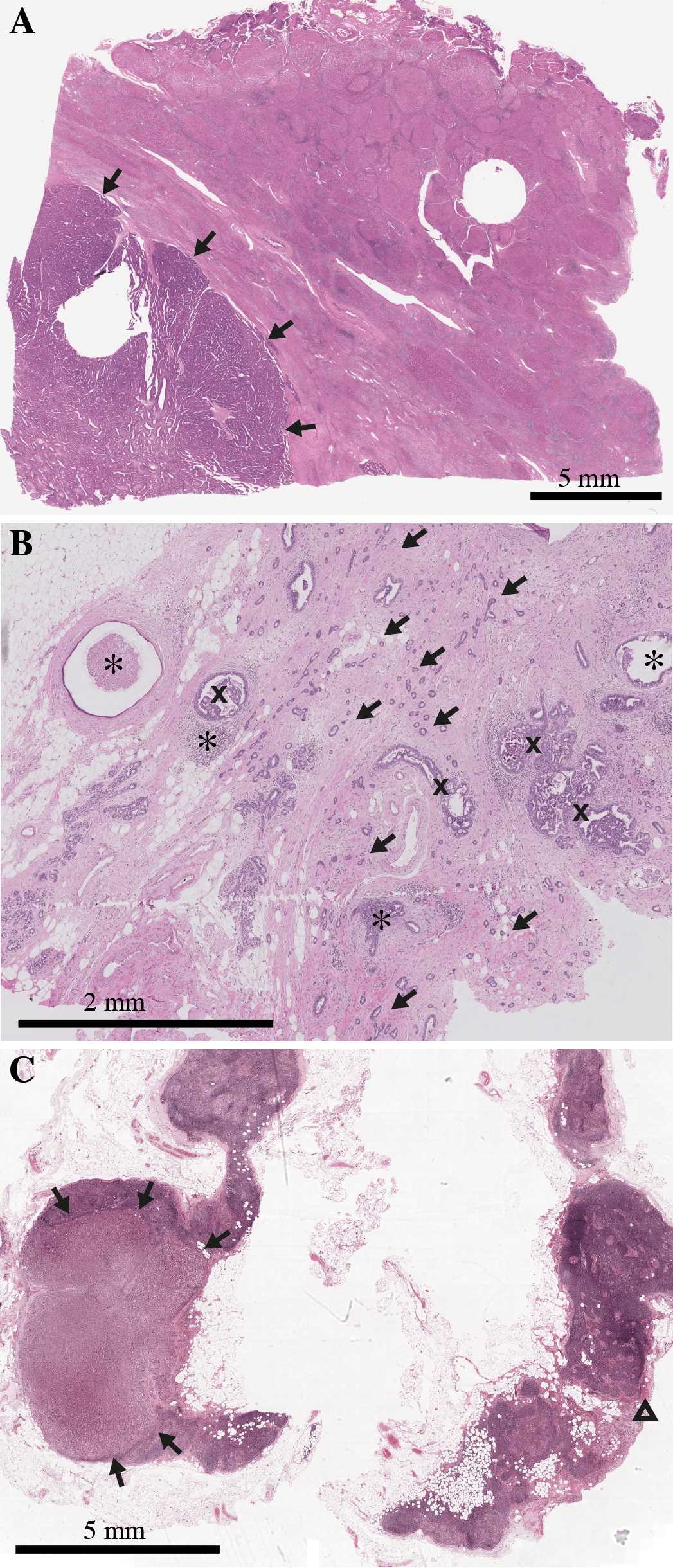}
    \caption{Exemplary cases from the PAIP 2019 (A), BACH 2018 (B) and CAMELYON 2016 (C) datasets. The arrows mark regions of hepatocallular carcinoma, breast carcinoma and a lymph node macrometastasis in A, B and C, respectively. In B, benign and in situ lesions are indicated by stars and x's. Beside the macrometastasis in C, also isolated tumor cells are found (next to the triangle).}
    \label{fig:dataset_example}
\end{figure}

Figure \ref{fig:dataset_example} (A) shows an exemplary whole-slide image from the dataset.
It illustrates why hepatocellular carcinoma (HCC) is a prominent example of a cancer that is characterized not only by nuclear abberations, but also by local tissue abnormalities and large-scale features such as a so-called pseudocapsule (as illustrated in the figure). In fact, (low-grade) HCC is challenging to diagnose, as it is often identifiable \textit{only} by abberations of the long-range tissue architecture with only minor nuclear abnormalities, if any \citep{WHODigestive}.
This makes the dataset well-suited for demonstration of the importance of multi- and particularly large-scale context features.

\subsubsection{BACH 2018}
\label{sec:BACH_description}
The BACH (BreAst Cancer Histology) 2018 challenge \citep{Aresta2019} included the classification of small ($2.048 \times 1.536$) image patches as one task, and the segmentation of WSIs of breast biopsies as another. For the latter task, 10 pixel-wise annotated WSIs were provided for training. Two medical experts performed the image segmentation using the following four labels: (1) normal, (2) benign, (3) in situ and (4) invasive breast carcinoma. WSIs were acquired by a Leica SCN400 ($\times 20$ power, pixel scale: \SI[per-mode=symbol]{0.467}{\micro\meter\per\pixel}) in the period from 2013 to 2015 at the Centro Hospitalar Cova da Beira. The image sizes range from $50,529 \times 36,833$ to $64,703 \times 45,808$ pixels (1.861 to 2.964 gigapixels).

Differential diagnosis in suspected breast cancer is known as a challenging task to human pathologists, and also to machine learning approaches. This is reflected in the results of the top-performing teams, reaching only moderate scores of $0.50$ to $0.69$ with respect to the custom challenge metric \citep{Aresta2019}.
Figure \ref{fig:dataset_example} (B) illustrates how close the different diagnoses can be interwoven, with invasive and in situ parts of the carcinoma directly neighboring each other.

\subsubsection{CAMELYON 2016/MM subset}
\label{sec:CAMELYON_description}
For the original CAMELYON (Cancer Metastases in Lymph Nodes) 2016 challenge \citep{EhteshamiBejnordi2017}, 399 WSIs of lymph nodes from women with confirmed breast cancer were gathered from two different centers (Radboud UMC and UMC Utrecht) during the first half of 2015. 240 of the 399 slides contained one or more nodal metastases.
The images were scanned by a Pannoramic 250 Flash II ($\times 20$, pixel scale: \SI[per-mode=symbol]{0.243}{\micro\meter\per\pixel}) and a NanoZoomer-XR Digital slide scanner C12000-01 ($\times 40$, pixel scale: \SI[per-mode=symbol]{0.226}{\micro\meter\per\pixel}), respectively. Image sizes are $61,440 \times 53,760$ to $217,088 \times 103,936$ pixels (3.303 to 22.563 gigapixels).

Initial annotations of lymph node metastases were drawn by medical students and then reviewed and corrected by two expert pathologists, as detailed in \citep{EhteshamiBejnordi2017}.
This procedure fits the clinical observation that the detection of lymph node metastasis is a much easier task to human pathologists than diagnosis of HCC and pathologies of the breast.
The original dataset contains macrometastases (tumor cell clusters with a diameter $\geq \SI{2}{\milli\meter}$), micrometastases ($\SI{0.2}{\milli\meter} \leq$ diameter $< \SI{2}{\milli\meter}$) as well as yet smaller clusters down to isolated tumor cells. Whilst these are all clinically relevant, isolated tumor cells and very small clusters do not infer with the global lymph node architecture \citep{WHOBreast}.

For our experiments, we therefore created a subset of the CAMELYON 2016 dataset, consisting of 20 WSIs with at least one macrometastasis. We refer to this subset as CAMELYON 2016/MM, with MM for macrometastasis. The procedure for the establishment of the CAMELYON 16/MM subset is detailed in section \ref{sec:supp_camelyon_subset}.
Figure \ref{fig:dataset_example} (C) provides a typical example of the CAMELYON 2016/MM dataset. As illustrated in figure \ref{fig:dataset_example} (C), this dataset still contains isolated tumor cells (triangle) and micrometastases along with the macrometastases (arrows), but with a higher weight put to the macrometastases and the larger length scales as compared to the original CAMELYON 2016 dataset. However, it should be noted that this task still differs from the other two, namely diagnosis of HCC and pathologies of the breast. Whilst macrometastases infer with the global architecture of the lymph node, they can still be diagnosed at the single- or few-cell level, by virtue of the differences of individual tumor and autochthonous cells. Therefore, even though larger scales certainly help the human pathologist with the diagnosis of macrometastases, the consideration of larger scales and the integration of multi-scale information is of lesser importance.

\subsection{Preprocessing}

WSI data preprocessing was similarly performed for all three datasets. Compared to the annotations originally provided for the respective challenges, we automatically generated annotations for an "overall tissue" as an additional class. This was achieved by thresholding of the original images with [R, G, B] $\leq$ [235, 210, 235], followed by binary morphological opening and closing operations.
The rationale for introducing the additional class was threefold: First, it makes the "background" class, that would otherwise include not only all-white background \textit{but also} healthy tissue, much less heterogeneous. Second, it facilitates sampling of healthy tissue patches as meaningful negative examples (cf. section \ref{sec:experiments}). %
Third, the automatically generated "overall tissue" annotations let us generalize the Reinhard color normalization \citep{Reinhard2001} to our application. Reinhard color normalization, as originally introduced, implicitly assumes the color statistics to be computed from meaningful image areas. In contrast, when directly applied to a WSI image, the fraction of all-white background would determine the color statistics and hence a color normalization, which is undesired. To avoid this, images were standardized by Reinhard colour normalization \citep{Reinhard2001} with respect to the "overall tissue" regions only, and then normalized to channel-wise zero average and unit variance, again in the tissue regions only.

\subsection{Evaluation metrics}
\label{sec:validation}

The primary outcome parameter for all experiments was the Jaccard index with the classes weighted as detailed in section \ref{sec:experiments}. Secondary, for experiments on the BACH challenge data, we additionally report the custom metric used in that specific challenge \citep{Aresta2019}. Simply put, the metric is based on a pixel-wise accuracy measure that penalizes predictions that are farther from the given ground truth segmentation class. This is possible since the segmentation classes can be ordered according to their malignancy. We refer to that metric as the ``BACH metric''.

For each dataset, we employed a 5-fold cross validation (CV) strategy with the split conducted at the level of the entire WSIs. Splitting is performed such that all classes were present in the validation set. The splits were computed once and then kept fixed throughout all experiments. The number of WSIs in the validation set was 10 for PAIP 2019, 2 for BACH 2018 and 4 for CAMELYON 2016/MM.
Validation was performed after the following epochs (number of iterations):
1 ($1,920$), 3 ($5,760$), 5 ($9,600$), 8 ($15,360$), 11 ($21,120$), 16 ($30,720$), 21 ($40,320$), and then every 10 epochs ($19,200$ iterations).
At each validation step, a fixed number of $3,072 \times 3,072$ pixel-sized sub-images per image was evaluated. We evaluated four such sub-images per validation WSI for PAIP, ten for BACH and six for CAMELYON 2016/MM, resulting in 40, 40 and 24 of these per spilt for PAIP, BACH and CAMELYON 2016/MM, respectively.
The positions of the sub-images were randomly sampled with the condition that, for any WSI, all available classes shall be represented in at least $N_\mathrm{I} / N_\mathrm{c}$ of the sub-images. Here, $N_\mathrm{I}$ denotes the number of validation sub-images from that WSI and $N_\mathrm{c}$ is the number of available classes. Sampling of the sub-image positions was done once, before the first experiment, and then kept fixed throughout all experiments. This means that all models for a given split are evaluated with respect to the exact same sub-images at all validation steps.

\subsection{Statistical analysis}
\label{sec:statistics}

For a given model, performance scores were evaluated per split, from which the mean and 95\%-confidence interval (CI) across the five splits were computed.
To set a robust baseline for the experiments in section \ref{subsec:Res1}, the experiments for the baseline, single-scale U-Net were repeated three times and averaged, corresponding to a conservative estimate (over clustered experiments). The results for the individual runs of the baseline U-Net are provided in the section \ref{sec:supp_baseline_results}.

As detailed in section \ref{sec:supp_statistical_significance_and_power}, we applied a corrected t-test for use in cross-validation settings \citep{NadeauBengio, BouckaertFrank} to test for statistical significance of the differences between the multi-scale models and the baseline U-Net. As further described in the same section, our study is underpowered for examining differences between different multi-scale models for significance. Therefore, we restricted the tests to comparisons between multi-scale architectures and the baseline U-Net.
All statistical analyses were performed using R v4.0.0 \citep{RPackage}.

\section{Experiments}
\label{sec:experiments}

Our experiments were designed as follows: First, we explored the behaviour of the msY-family on the PAIP 2019 dataset. Second, to examine generalizability of our findings, we evaluated the performance of the best performing architectures on the BACH 2018 and the CAMELYON 2016/MM dataset. Third, to aetiologically understand the behaviour of the msY-models and to narrow down the possible design variants, we conducted further experiments with selected variants of that models on the PAIP 2019 dataset.

The training strategy and all hyperparameters were manually optimized for training of the baseline Res18-Unet on PAIP 2019 before the experiments and then kept fixed for all models and throughout all experiments.
We aimed to reproduce a standard and widely used setup as much as possible, including the use of a common model as the baseline.
For the same reason, we chose the binary cross entropy (BCE) loss as a common and widely used loss function (see, e.g., \citealt{heller2020kits}). BCE was used as a loss function for both classification (if the model has an additional classification output, cf. section \ref{subsec:context_gate}) and segmentation.
The training process was split into (short pseudo\hbox{-})epochs of $1,920$ patches each, with all patches re-sampled after each individual "epoch". Compared to the use of a fixed number of pre-selected patches, this reduces overfitting given the limited memory resources and the large whole-slide images.
For each pseudo-epoch, the patches were balanced with respect to both the individual cases in the training set and the available classes.
Optimization was performed using Adam \citep{kingma2014adam} employing a learning rate of $10^{-3}$ and a learning rate decay with $\gamma = 0.5$ every 30 epochs (57,600 iterations).
During training, we employed online data augmentation including the following standard operations: rotation, flip along the horizontal axis, saturation and brightness transformation.

\subsection{Comparison of msY-family and baseline models}

\subsubsection{PAIP 2019}
\label{subsec:Exp1}

The clinical scenario in our PAIP 2019 experiments is the segmentation of hepatocellular carcinoma for evaluation of tumor extent and margin status after the resection of hepatocellular carcinoma.

The comparison of all multi-scale models (including the msY-family and the multi-scale ensembling variants) to the baseline U-Net resulted in a total of $N=9$ pairwise model comparisons. The evaluation of the corresponding results were the primary focus of the PAIP 2019 experiments.
In addition, we also compared the different multi-scale architectures amongst each other and to multi-scale ensembles. However, as our study is underpowered for finding statistically significant results in between different multi-scale models (cf. section \ref{sec:supp_statistical_significance_and_power}), we only descriptively present the corresponding results.

For the PAIP 2019 experiments, all models were trained for at least 120 pseudo-epochs ($230,400$ iterations) or until convergence of the validation loss was reached. According to the clinical scenario and due to  the fact that the localisation of viable tumor cells with respect to the resection margins determines the resection status, we put additional weight to the viable tumor class as compared to the whole tumor class and weight the loss by [0, 1, 2, 6] for background, overall tissue, whole tumour and viable tumor classes, respectively.

Segmentation performance as the primary outcome parameter was measured by the weighted average Jaccard index in the classes of interest, i.e., ``whole tumor'' and ``viable tumor'', where the weights are the same as above.

As a secondary outcome, we furthermore examined the resource requirements of the models. We report the number of trainable parameters and measured the GPU memory footprint. For the latter, we observed the GPU memory usage of the training process, including forward and backward pass, at different batch sizes using the NVIDIA System Management Interface. The memory footprint of the model without system overhead was then deduced by linear regression and reported as gigabytes per patch in batch. The detailed measurements are reported in section \ref{sec:supp_memory_footprint_results}.
For multi-scale models, the term
``patch'' is meant as a multi-scale patch, i.e., including the image patches for all individual scales. The input patch sizes in all models and for all scales are $512 \times 512$ pixels, except at scale 4 ($576 \times 576$ for the U-Net in the ensembles, $572 \times 572$ for the msY-family models).

\subsubsection{BACH 2018}
\label{subsec:Exp1_BACH}

As described in section \ref{sec:BACH_description}, the BACH 2018 dataset allows us to evaluate the models with respect to their capability in the differential diagnoses of breast lesions, which represents the clinical scenario for this experiment.

We started from the models from the PAIP 2019 experiments and, after replacement of the output layers, fine-tuned them on the BACH 2018 dataset.
Without loss of generality, we fine-tuned the models from the split $i$ ($i \in [0, 1, 2, 3, 4]$) of the PAIP dataset on the split with the same index $i$ ($i \in [0, 1, 2, 3, 4]$) of the BACH 2018 dataset.
Training was performed exactly as described in section \ref{subsec:Exp1}, with the following differences: First, for the first 10 pseudo-epochs ($19,200$ iterations), only the output layers were trained and the other weights were kept fixed. Fine-tuning of all weights was then continued for another 70 pseudo-epochs ($134,400$ iterations).
Second, in line with the clinical scenario of breast lesion differential diagnosis, we put equal weights to all classes of interest, i.e., ``normal'', ``benign'', ``in situ'' and ``invasive carcinoma'', both for the computation of the loss function and the weighted-average Jaccard index.
In order to examine whether our models are on par with state-of-the-art results on this task, we additionally evaluated the scores from the custom BACH 2018 metric as introduced in \citep{Aresta2019}.

\subsubsection{CAMELYON 2016/MM}
\label{subsec:Exp1_Camelyon}
For our CAMELYON 2016/MM experiments, we again started using the models from the PAIP 2019 experiments and, after replacement of the output layers, fine-tuned them on the new dataset. Training was again performed as described in section \ref{subsec:Exp1} except that for the 10 pseudo-epochs ($19,200$ iterations) only the output layers were trained and the other weights were kept fixed. For CAMELYON 2016/MM dataset, we fine-tuned for at least 60 pseudo-epochs ($115,200$ iterations) or until convergence. Again, without loss of generality, we fine-tuned the models from the split $i$ ($i \in [0, 1, 2, 3, 4]$) of the PAIP 2019 dataset on the split with the same index $i$ ($i \in [0, 1, 2, 3, 4]$) of the CAMELYON 2016/MM dataset.

\subsection{Context classification loss}
\label{subsec:Exp3}

We hypothesized that when using a global-scale encoder, introduction of an additional context  classification loss as a means of "guidance" may benefit the model (cf. section \ref{subsec:context_loss}). In order to examine whether this is consistently the case, we trained the following models with and without context classification loss on the PAIP 2019 dataset: msY$_{(16)}$, msY$^2$, msY$_{(16),\text{MM}}$.

\subsection{Spatial alignment in multi-scale merge blocks}
\label{subsec:Exp2}

We hypothesized that spatial matching is an essential step in the multi-scale merge block. To test this hypothesis, we compared a msY$^2$-Net, as an example where three different scales are merged, to a variant of the same model but with the multi-scale merge block replaced by a pure concatenation followed by a $1 \times 1$ convolution. This corresponds to a multi-scale merge block but without alignment of the spatial scales and without preserving spatial relationships. The rest of the spatially non-aligned msY$^2$-Net
variant remained unchanged.

If, in line with the hypothesis, the spatially non-aligned msY$^2$-Net performs worse than its standard variant with the correct multi-scale merge block, it remains to be examined whether this is "only" due to the clumsy initialization and can potentially be overcome without spatial merging. One might hypothesize that the deficit
is not architecturally, but due to the introduction of new, untrained convolutions in the middle of the otherwise pretrained encoder. Therefore, we examined two different variants of the non-aligned model:
one variant with randomly initialized weights
and another variant
where the $1 \times 1$ convolution was initialized with the unit matrix at the channels belonging to the main encoder and with 0's everywhere else, both superimposed with random noise (normal distribution with standard deviation $10^{-4}$). In the latter variant, the main encoder corresponded to the unperturbed, pretrained model at the beginning of the training process, up to noise.

\subsection{Multiple merging}
\label{subsec:Exp4}

The multi-scale merge block as presented in section \ref{subfig:msM-Block} can be introduced at any level of the encoders. Therefore, it also allows for earlier or multiple fusions, which may allow for the additional processing of combined features.
Very early merge blocks, however, do not provide much additional context, due to the cropping step inside the multi-scale merge block.
Therefore, it is not a priori clear, at which level the merge connection shall be established or if multiple path merges can further benefit the model.

In order to study whether models with multiple merges can be trained robustly and whether an effect through multiple merges can be found, we compared the segmentation performance of a msY$_{(16)}$-, a msY$^2$- and a msY\textit{I}-Net to the analogues of them with multi-scale merge blocks at all encoder levels. In this experiment, to examine a "maximum" msY-family variant trainable on our hardware, we used ResNet34 instead of a ResNet18 for the global context encoder of the msY$^2$-model. All other models and encoders were ResNet 18-based, as by our standard. Additionally, we examined to which extent the introduction of multiple merges increases the GPU memory footprint of the models.

\section{Results}
\label{sec:results}

\subsection{Multi-scale multi-encoder models improve histopathology image segmentation}
\label{subsec:Res1}

Table \ref{tab:exp1_paip} reports the weighted average Jaccard index for the two non-trivial classes of the PAIP 2019 dataset, viable tumor and whole tumor (including stroma etc.). According to the clinical scenario of evaluating tumor extent and resection margin status in hepatocelullar carcinoma, 3x more weight is put onto the viable tumor class than on the whole tumor class (cf. section \ref{subsec:Exp1} for details).

\begin{table*}[]
    \caption{Different models from the proposed multi-scale multi-encoder family and multi-scale ensembles versus the baseline U-Net. The figures in the table depict the class-weighted Jaccard index for whole and viable tumor classes on the PAIP 2019 dataset. A $\star$ denotes statistical significance at the level of $0.05$ when compared to the scale 1-U-Net.
    	For a quick overview, the best results per split and overall are marked in bold, ignoring differences $<0.005$.
    	}
    \centering\small
    \begin{threeparttable}
    \begin{tabular}{p{1.285cm} p{0.85cm} p{0.85cm} p{0.7cm} P P P P >{\scriptsize}p{0.4cm} p{2.6cm} >{\scriptsize}p{0.05cm}}
    \toprule
    \textbf{Arch.} & \textbf{Scales} & \textbf{\#\,pms.\tnote{$\dagger$}} & \textbf{Mem.\tnote{$\ddagger$}} & \multicolumn{6}{c}{\textbf{Weighted Jaccard}}  & \\
    \cmidrule(r){5-10}
    \addlinespace
     &   &   &   &   \multicolumn{5}{c}{\textbf{Per CV fold}}  & \textbf{Mean (95\% CI)}  \\
    \hline
\addlinespace[\betweenmodels]
U-Net & 1 & 17.804 & 1.066 & 0.814 & 0.754 & 0.686 & 0.735 & 0.729
 & 0.744 (0.707, 0.780)
 &
\\
\addlinespace[\betweenmodels]
Avg. Ens.& 1,\,4,\,16 & 53.412\tnote{\S} & 3.506\tnote{\S} & 0.890 & 0.853 & \textbf{0.836} & 0.900 & 0.835
 & \textbf{0.863 (0.839, 0.887)}
 & $\star$\\
Log. Ens.& 1,\,4,\,16 & 53.412\tnote{\S} & 3.506\tnote{\S} & 0.859 & 0.814 & 0.802 & 0.915 & 0.818
 & 0.842 (0.805, 0.878)
 & $\star$\\
Maj. Ens.& 1,\,4,\,16 & 53.412\tnote{\S} & 3.506\tnote{\S} & 0.883 & 0.838 & 0.770 & 0.875 & 0.816
 & 0.836 (0.800, 0.873)
 & $\star$\\
\addlinespace[\betweenmodels]
\multirow{2}{*}{msY-Net}
& 1,\,4 & 29.282 & 1.422 & \textbf{0.902} & 0.800 & 0.712 & 0.796 & 0.794
 & 0.801 (0.748, 0.854)
 & $\star$\\
     & 1,\,16 & 29.284& 1.175 & 0.861 & 0.872 & 0.789 & 0.873 & 0.889
 & 0.857 (0.826, 0.887)
 & $\star$\\
\addlinespace[\betweenmodels]
\multirow{2}{*}{msU\textit{I}-Net}
       & 1,\,4 & 28.983 & 1.389 & 0.833 & 0.811 & 0.779 & 0.858 & 0.898
 & 0.836 (0.800, 0.871)
 & $\star$\\
      & 1,\,16 & 28.983 & 1.355 & 0.864 & 0.841 & 0.785 & 0.896 & 0.889
 & 0.855 (0.820, 0.890)
 & $\star$\\
\addlinespace[\betweenmodels]
msY\textit{I}-Net& 1,\,4,\,16 & 40.460 & 1.398 &  0.847 & \textbf{0.871} & 0.804 & 0.895 & \textbf{0.910}
 &\textbf{0.865 (0.833, 0.898)}
 & $\star$\\
\addlinespace[\betweenmodels]
msY$^2$-Net& 1,\,4,\,16 & 40.761 & 1.561 & 0.847 & \textbf{0.873} & 0.795 & \textbf{0.934} & 0.876
 & \textbf{0.865 (0.825, 0.905)}
 & $\star$\\
\\
    \bottomrule
    \end{tabular}
    \begin{tablenotes}
    \item[\S] sum of three individual U-Nets (scales: 1, 4, 16)
    \item[$\dagger$] in units of one million parameters
    \item[$\ddagger$] GPU memory footprint given in gigabytes per image patch in batch
    \end{tablenotes}
    \end{threeparttable}
    \label{tab:exp1_paip}
\end{table*}

\begin{table*}[]
    \caption{The best performing multi-scale architectures versus the baseline U-Net on the BACH 2018 dataset. The figures in the table depict the class-average of the Jaccard index for normal tissue, benign lesions and in situ and invasive carcinoma. A $\star$ denotes statistical significance at the level of $0.05$ when compared to the scale 1-U-Net.
    	The best results per split and overall are marked in bold, ignoring differences $<0.005$.
    	}
    \centering\small
    \begin{threeparttable}
    \begin{tabular}{p{1.285cm} p{0.8cm} p{0.8cm} p{0.67cm} P P P P >{\scriptsize}p{0.38cm} p{2.55cm} >{\scriptsize}c}
    \toprule
    \textbf{Arch.} & \textbf{Scales} & \textbf{\#\,pms.\tnote{$\dagger$}} & \textbf{Mem.\tnote{$\ddagger$}} & \multicolumn{6}{c}{\textbf{Weighted Jaccard}}  & \\
    \cmidrule(r){5-10}
    \addlinespace
     &   &   &   &   \multicolumn{5}{c}{\textbf{Per CV fold}}  & \textbf{Mean (95\% CI)}  \\
    \hline
\addlinespace[\betweenmodels]
U-Net & 1 & 17.804 & 1.066 & 0.523 & 0.476 & 0.478 & 0.510 & 0.501
 & 0.498 (0.482, 0.514)
 \\
\addlinespace[\betweenmodels]
Avg. Ens. & 1,\,4,\,16 & 53.412\tnote{\S} & 3.506\tnote{\S} & 0.593 & 0.540 & 0.518 & 0.581 & \textbf{0.622}
 & 0.571 (0.538, 0.603) & $\star$
\\
\addlinespace[\betweenmodels]
msY\textit{I}-Net
& 1,\,4,\,16 & 40.460 & 1.398 &  \textbf{0.629} & 0.569 & 0.488 & 0.633 & 0.553
 & \textbf{0.574 (0.527, 0.621)} & $\star$
\\
\addlinespace[\betweenmodels]
msY$^2$-Net & 1,\,4,\,16 & 40.761 & 1.561 & 0.578 & \textbf{0.594} & \textbf{0.536} & \textbf{0.662} & 0.524
 & \textbf{0.579 (0.536, 0.622)} & $\star$
\\
    \bottomrule
    \end{tabular}
    \begin{tablenotes}
    \item[\S] sum of three individual U-Nets (scales: 1, 4, 16)
    \item[$\dagger$] in units of one million parameters
    \item[$\ddagger$] GPU memory footprint given in gigabytes per image patch in batch
    \end{tablenotes}
    \end{threeparttable}
    \label{tab:exp1_bach}
\end{table*}

\begin{table*}[]
    \caption{The best performing multi-scale architectures versus the baseline U-Net on the CAMELYON 2016/MM dataset. A ($\star$) denotes statistical significance at the level of $0.05$ when individually compared to the scale 1-U-Net (not correcting for multiple testing).
    For the comparisons of other models to the baseline U-Net, the null hypothesis cannot be rejected at $0.05$.
    	The best results per split and overall are marked in bold, ignoring differences $<0.005$.
    	}
    \centering\small
    \begin{threeparttable}
    \begin{tabular}{p{1.285cm} p{0.85cm} p{0.85cm} p{0.7cm} P P P P >{\scriptsize}p{0.4cm} p{2.6cm} >{\scriptsize}p{0.05cm}}
    \toprule
    \textbf{Arch.} & \textbf{Scales} & \textbf{\#\,pms.\tnote{$\dagger$}} & \textbf{Mem.\tnote{$\ddagger$}} & \multicolumn{6}{c}{\textbf{Jaccard (lymph node metastases)}}  & \\
    \cmidrule(r){5-10}
    \addlinespace
     &   &   &   &   \multicolumn{5}{c}{\textbf{Per CV fold}}  & \textbf{Mean (95\% CI)}  \\
    \hline
\addlinespace[\betweenmodels]
U-Net & 1 & 17.804 & 1.066 & 0.696 & 0.770 & 0.805 & 0.903 & 0.803
 & 0.796 (0.737, 0.854)
 &
 \\
\addlinespace[\betweenmodels]
Avg. Ens. & 1,\,4,\,16 & 53.412\tnote{\S} & 3.506\tnote{\S} & 0.748 & \textbf{0.944} & 0.798 & 0.923 & 0.883 & \textbf{0.859 (0.794, 0.924)}
\\
\addlinespace[\betweenmodels]
msY\textit{I}-Net & 1,\,4,\,16 & 40.460 & 1.398 & 0.710 & 0.756 & 0.822 & 0.874 & 0.859 & 0.804 (0.750, 0.859)
\\
\addlinespace[\betweenmodels]
msY$^2$-Net & 1,\,4,\,16 & 40.761 & 1.561 & \textbf{0.775} & 0.804 & \textbf{0.865} & 0.918 & \textbf{0.889}
 & 0.850 (0.804, 0.897)
 & ($\star$)
\\
    \bottomrule
    \end{tabular}
    \begin{tablenotes}
    \item[\S] sum of three individual U-Nets (scales: 1, 4, 16)
    \item[$\dagger$] in units of one million parameters
    \item[$\ddagger$] GPU memory footprint given in gigabytes per image patch in batch
    \end{tablenotes}
    \end{threeparttable}
    \label{tab:exp1_camelyon}
\end{table*}

The first experiment shows a number of aspects: First and most importantly, there is a considerable improvement over the baseline U-Net by adding multi-scale input, either through ensembling or by multiple encoders as in the msY-family.
The effect is found to be statistically significant in our experiments, even when tested only for five CV folds and even with a conservative correction for both the multiplicity of the pairwise comparisons and the violation of the independent samples-assumption underlying standard t-statistics.

The proposed multi-scale multi-encoder models reach the same segmentation performance as our best multi-scale ensemble, but as individual end-to-end trainable models and with much lower resource requirements. Taking $msY^2$-Net as an example, it has only $76.3 \%$ of the parameters of a the corresponding ensemble of three U-Nets and comes with a GPU footprint reduced by $54.5 \%$ (if trained in parallel).
Concerning the pairwise comparisons between different msY family architectures, our data suggest that, for the PAIP 2019 dataset, the global context patch provides more valuable information than the local context patch.
However, the combination of both local and global context appears to lead to a further improvement, as the msY$^2$- and the msY\textit{I}-Net models are consistently found amongst the top performing approaches, but this effect seems marginal and cannot be reliably detected through this study.

We next examined whether the superior performance of the multi-scale architectures translates to other datasets, tumor entities and tasks. For the task of breast lesion segmentation and differential diagnosis, the class-wise mean Jaccard indices are reported in table \ref{tab:exp1_bach}. The results are in line with our findings on the PAIP 2019 dataset.
Notably, the Jaccard indices on the BACH 2018 dataset are globally lower than on the PAIP 2019 dataset for all architectures. This observation fits the considerations in section \ref{sec:BACH_description} that this is a particularly challenging task, also for the human pathologist.
In addition, table \ref{tab:supp_exp1_bachmetric} reports the performances of the same architectures as evaluated by the custom Bach metric used in the original challenge (cf. \citep{Aresta2019} for details). The metric values for all architectures discussed in this section are in the range of the challenge results, with the multi-scale architectures, including the msY-family models and the multi-scale ensembles, reaching top performance.

Finally, table \ref{tab:exp1_camelyon} shows the performance of corresponding models applied to lymph node metastases segmentation. The results are again in line with the findings on PAIP 2019 and BACH 2018 data. However, the performance increase through the use of multiple scales is smaller in CAMELYON 2016/MM (6.8\% as compared to 16.3\% for both PAIP 2019 and BACH 2018), which fits our considerations on the different nature and difficulty  of these tasks (cf. section \ref{sec:CAMELYON_description}).

With respect to the generalizability of these results, we further note that, later, an independent group has been able to present additional evidence in favor of these findings on yet other data \citep{van_rijthoven_hooknet_2020}. The group used a related but different end-to-end-trainable multi-scale model that confirms the benefit from the introduction of a context-encoder, though they train their context-encoder using an additional decoder that makes the model much heavier in terms of trainable parameters and GPU usage and voids the resource improvements as compared to multi-scale ensembles.

\subsection{Context classification loss-based deep guidance for context encoder training}
\label{subsec:Res4}

In order to examine whether the additional classification loss is indeed helpful for ``guidance'' of the global context encoder during training, we re-examined the msY$^2$-, the msY$_{(16)}$- and the msY\textit{I}-Net models on the PAIP 2019 dataset when trained with or without the additional classification loss.

From the results in table \ref{tab:exp2_class_loss}, it appears that the additional use of the classification loss consistently improves the model training.
Recalling that the relative sizes of the detail and the global context encoder input patches (cf. figure \ref{fig:patches}) translate to a corresponding center-crop operation on the context encoder feature maps (as part of the multi-scale merge block, cf. figure \ref{subfig:msM-Block}), this behaviour might be understood intuitively as the the classification loss helping the context encoder acquire additional gradients for training.

Importantly, this technique comes with a very moderate overhead both in terms of the number of additional parameters and the effective GPU memory footprint.

\begin{table*}[]
    \caption{Context classification loss guidance for training of the global context encoder.
    Numbers are the class-weighted Jaccard indices for the PAIP 2019 dataset.
    The best results per pairwise comparison and split are marked in bold, where differences $<0.005$ are ignored.}

        \centering\small
    \begin{threeparttable}
    \begin{tabular}{p{1.285cm} p{1.3cm} p{0.85cm} p{0.7cm} P P P P >{\scriptsize}p{0.4cm} p{2.6cm}}
    \toprule
    \textbf{Arch.} & \textbf{Clss.\,loss} & \textbf{\#\,pms.\tnote{$\dagger$}} & \textbf{Mem.\tnote{$\ddagger$}} & \multicolumn{6}{c}{\textbf{Weighted Jaccard}} \\
    \cmidrule(r){5-10}
    \addlinespace
     &  & & & \multicolumn{5}{c}{\textbf{Per CV fold}}  & \textbf{Mean (95\% CI)}  \\
    \hline
    \addlinespace[\betweenmodels]
\multirow{2}{*}{msY$^2$-Net} & with & 40.761 & 1.561 & 0.847 & \textbf{0.873} & \textbf{0.795} & \textbf{0.934} & \textbf{0.876}
 & \textbf{0.865 (0.825, 0.905)}\\
     & without & 40.759 & 1.436 & \textbf{0.903} & 0.851 & 0.737 & 0.826 & 0.779
 & 0.819 (0.769, 0.869)\\
\addlinespace[\betweenmodels]
\multirow{2}{*}{msY$_{16}$-Net} & with & 29.284 & 1.281 & 0.861 & \textbf{0.872} & \textbf{0.789} & \textbf{0.873} & \textbf{0.889}
 & \textbf{0.857 (0.826, 0.887)}\\
 & without & 29.282 & 1.175 & \textbf{0.883} & 0.847 & 0.713 & 0.715 & 0.764
 & 0.785 (0.724, 0.845)\\
\bottomrule
    \end{tabular}
    \begin{tablenotes}
    \item[$\dagger$] in units of one million parameters
    \item[$\ddagger$] GPU memory footprint given in gigabytes per image patch in batch
    \end{tablenotes}
    \end{threeparttable}
    \label{tab:exp2_class_loss}
\end{table*}

\subsection{Necessity for spatial alignment in multi-scale merge blocks}
\label{subsec:Res2}

Drawing on theoretical considerations, we have constructed the multi-scale merge blocks such that spatial relationships between the different paths are preserved upon fusion.
Table \ref{tab:exp3_avg_bestloss} repeats the results for the msY$^2$-Net and compares them to those of two variants of the same architecture which both do \textit{not} adhere to this condition. These are constructed by concatenating the bottleneck feature maps from the different paths without any spatial alignment as part of the multi-scale merge block, where the two variants differ in that the merging $1 \times 1$ convolution is either initialized randomly
or with pre-defined weights leaving the main path unperturbed at training startup
(cf. section \ref{subsec:Exp3} for details).

It can be seen that both these variants perform consistently worse than the proposed msY$^2$-Net with correct spatial alignment in the multi-scale merge block.
This is irrespective of whether the weights of the $1 \times 1$ convolution are randomly initialized or whether they are initialized such that the ResNet18 encoder of the underlying U-Net is left untouched by the un-aligned merge connections.
This suggests that the deficit through the missing alignment step may be architectural rather than only a disturbance of the pretrained encoder by the additional $1 \times 1$ convolution.
We therefore conclude that, when merging paths at the bottleneck level in a manner as done by the multi-scale merge block, spatial matching is a necessary step.

\begin{table*}[]
    \caption{Spatial alignment at multi-scale path fusion. Numbers are the class-weighted Jaccard indices for the PAIP 2019 dataset.
    All models are variants of the msY$^2$ architecture, either with spatial alignment at the multi-scale merge block (1, as per default) or without (2, 3). The best results per pairwise comparison and split are marked in bold, where differences $<0.005$ are ignored.}
    \centering\small
    \begin{tabular}{p{3.5cm} P P P P >{\footnotesize}p{1.0cm} p{2.6cm}}
    \toprule
    \textbf{msY$^2$-Net variant} & \multicolumn{6}{c}{\textbf{Weighted Jaccard}} \\
    \cmidrule(r){2-7}
    \addlinespace
     &  \multicolumn{5}{c}{\textbf{Per CV fold}}  & \textbf{Mean (95\% CI)}  \\
    \hline
    \addlinespace[\betweenmodels]
spatially aligned$^1$ & \textbf{0.870} & \textbf{0.835} & \textbf{0.796} & \textbf{0.902} & \textbf{0.865}
 & \textbf{0.854 (0.822, 0.885)}
 \\
 \addlinespace[\betweenmodels]
non-aligned, init$^2$  & \textbf{0.872} & 0.814 & 0.725 & 0.822 & 0.827
 & 0.812 (0.770, 0.854)
 \\
 \addlinespace[\betweenmodels]
non-aligned, random$^3$  & \textbf{0.875} & 0.787 & 0.777 & 0.808 & 0.832
 & 0.816 (0.785, 0.847) \\

    \bottomrule
    \end{tabular}
    \label{tab:exp3_avg_bestloss}
\end{table*}

\subsection{Additional path fusions can be added by introduction of multiple multi-scale merge blocks}
\label{subsec:Res3}

Table \ref{tab:exp4_avg_bestloss} compares the standard msY$^2$-, the msY$_{(16)}$- and the msY\textit{I}-Net with multi-scale merge blocks only at the bottleneck to their respective analogues with multi-scale merge connections at every level of the encoder.

\begin{table*}[]
    \caption{Multi-level multi-scale merging: For three different multi-scale architectures, the standard variant with multi-scale fusion by a single multi-scale merge block only at the bottleneck ("bottleneck") is compared to an analogous architecture, but with multi-scale merge blocks at each individual encoder level ("multiple").
    Numbers are the class-weighted Jaccard indices at the PAIP 2019 dataset.
    The best results per pairwise comparison and split are marked in bold, where differences $<0.005$ are ignored.}

    \centering\small
    \begin{threeparttable}
    \begin{tabular}{p{1.6cm} p{1.2cm} p{0.72cm} p{0.65cm} P P P P >{\scriptsize}p{0.38cm} p{2.55cm}}
    \toprule
    \textbf{msY$^2$-Net variant} & \textbf{Merge block(s)} & \textbf{\#\,pms.\tnote{$\dagger$}} & \textbf{Mem.\tnote{$\ddagger$}} & \multicolumn{6}{c}{\textbf{Weighted Jaccard}} \\
    \cmidrule(r){5-10}
    \addlinespace
     &  &  & & \multicolumn{5}{c}{\textbf{Per CV fold}}  & \textbf{Mean (95\% CI)}  \\
    \hline
    \addlinespace[\betweenmodels]

\multirow{2}{*}{msY$_{(16)}$-Net} & bottleneck & 29.284 & 1.175 & \textbf{0.861} & \textbf{0.872} & \textbf{0.789} & 0.873 & 0.889
 & \textbf{0.857 (0.826, 0.887)}
 \\
 & multiple & 29.464 & 1.281 & 0.854 & 0.848 & 0.769 & \textbf{0.881} & \textbf{0.903}
 & 0.851 (0.811, 0.891)
 \\
\addlinespace[\betweenmodels]
\multirow{2}{*}{msY\textit{I}-Net} & bottleneck & 40.460 & 1.398 & 0.847 & \textbf{0.871} & \textbf{0.804} & 0.895 & \textbf{0.910}
 & \textbf{0.865 (0.833, 0.898)}
 \\
 & multiple & 41.032 & 1.556 & \textbf{0.859} & \textbf{0.872} & \textbf{0.805} & \textbf{0.911} & 0.897
 & \textbf{0.869 (0.836, 0.901)}
 \\
\addlinespace[\betweenmodels]
\multirow{2}{*}{msY$^2_\text{(Res34)}$-Net} & bottleneck & 50.869 & 1.516 & 0.870 & 0.835 & 0.796 & \textbf{0.902} & 0.865
 & 0.854 (0.822, 0.885)
 \\
     & multiple & 51.140 & 1.608 & \textbf{0.876} & \textbf{0.879} & \textbf{0.802} & 0.863 & \textbf{0.915}
 & \textbf{0.867 (0.835, 0.899)}
 \\
\bottomrule
    \end{tabular}
    \begin{tablenotes}
    \item[$\dagger$] in units of one million parameters
    \item[$\ddagger$] GPU memory footprint given in gigabytes per image patch in batch
    \end{tablenotes}
    \end{threeparttable}
    \label{tab:exp4_avg_bestloss}
\end{table*}

Our study cannot find relevant differences between single-merge and multiple-merge setups. However, the additional merge connections did at least not lead to any instability in model training. Moreover, they come with a moderate increase in the number of trainable parameters and GPU memory requirements only. Therefore, multiple path fusions can be readily included in msY family models and can be used for further optimization, e.g. by systematic neural architecture searches.

\section{Discussion and Conclusions}
\label{sec:conclusion}

Using the segmentation of
carcinoma in hematoxylin-eosin (H\&E) stained whole-slide images %
as an example task,
our results show that the extensive integration of widely different spatial scales, as a ``mimicry'' of how humans approach analogous tasks, offers significant and relevant improvements over baseline single-scale U-Nets as the de facto standard in histopathology image segmentation.
The improvement has been consistently shown for three different datasets, clinical scenarios and tumor entities.

From a methodical and architectural perspective, our study presents a family of models that can integrate context from multiple scales and at various levels.
As an overarching effect, it shows that when fusing encoder paths from different scales, spatial alignment and the preservation of spatial relationships is necessary. The proposed multi-scale merge block fulfils this requirement.
The performance of the proposed architectures in terms of segmentation accuracy was further shown to be (at least) on par with ensembles of U-Nets, where the proposed multi-scale models models are end-to-end trainable systems with a reduced number of parameters and smaller memory footprint.
It should, however, be noted that only common ensembling techniques have been studied, and more sophisticated techniques could potentially lead to further improvement. Moreover, different to \citet{karimi2020DLGleason}, the logistic regression-based ensembling did not lead to better performance than standard averaging of the class probabilities of the U-Net trained for the different scales. The reason remains so far unclear; potential explanations could be the different application addressed (Gleason scoring vs. tumor segmentation) and differences in implementation.

It goes without saying that future research is also likely to optimize the multi-scale models presented herein much further.
As the detailed structure of encoder, decoder and possible skip connections are left entirely untouched, the proposed multi-scale architectures can seamlessly be adopted to various encoder-decoder models.
In particular, increased receptive fields in the individual encoders may benefit msY family models, as these would allow the center-crop of multi-scale merge block to acquire high-level features from a larger region.
Therefore, the use of dilated or atrous convolutions and atrous spatial pyramid pooling \citep{Chen2017DeepLab, Chen2017Atrous} in the context encoders might considerably benefit the models of the proposed multi-scale architecture family.
Continuing with potential for improvement, we note that -- as a straightforward extension of standard single scale CNN-based segmentation -- we trained our networks and encoders on the original "raw" image information. This approach neglects classical work on multi-resolution image representation, which could also be advantageous in the given application context.
Moreover, adding multi-level wavelet transforms to the CNN architecture has recently be shown as a very promising means to enlarge perceptive fields  \citep{liu_multi-level_2018,savareh_wavelet-enhanced_2019}, which, as said, can be particularly beneficial when implemented in the context encoders.

We have shown how the proposed building blocks and the underlying intuition can be extended to any number of arbitrarily sized spatial scales, as relevant for the particular organ and disease of interest.
Moreover, additional and early merge connections may offer the possibility to model more complex relations between the different scales.
We have shown that the proposed multi-scale merge block can be flexibly used at many levels, and that even msY family models with path fusions at all possible encoder levels can be trained robustly and with minor GPU memory overhead.
As both
the relevant spatial scales as well as the complexity of the tasks %
vary between applications, we finally envision that the msY family architectures with multiple merges open up a rich environment and search space also for neural architecture searches.

Apart from the pure network architecture perspective, we used binary cross entropy as a very common, widely used and accepted loss function for the current study. The results, however, already reveal that the integration of an additional context content loss (although again implemented as cross entropy) improves performance. This suggests that the application- and/or architecture-specific loss function design bears the potential for a further improvement of WSI segmentation performance.
Furthermore, we directly made use of the labels and annotations as provided by the respective challenge organizers for our experiments. It is, however, well known that the annotation of WSI slides is prone to errors and inter-observer variability. Over the past years, handling of label noise and uncertain annotation has attracted increasing interest and attention in the medical image analysis domain and in computational pathology, where, again, design of specific loss functions provides a promising approach (see \citet{karimi_deep_2020} for a recent review).

Concerning the limitations of this study, the results are so far based on three publicly available H\&E stained WSI image datasets; it remains to be seen in future work whether the results generalize to different image data, other tasks, including tumor grading and regression tasks (e.g. for survival prediction), and other diseases.
Furthermore, our analysis was
built on a cross validation strategy; therefore, confirmation of our findings on separate independent test databases would clearly be desirable.
Finally, whilst superior performance was consistently found for the three independent datasets, there still is a dataset dependency. In line with the underlying motivation of a human pathologist's mimicry, the data suggest that the performance increase is larger on tasks that require the human pathologist to integrate information from different scales (PAIP 2019, BACH 2018) and smaller on simpler tasks with more focus on individual cells (CAMELYON 2016/MM).

Despite the remaining limitations and potential for future work, the presented study provides clear evidence that a mimicry of how human experts approach a specific task can be successfully used to develop specialized machine learning architectures.
It advocates the integration of extensive multi-scale context into deep learning models for complex tasks in computational histopathology.

\section*{Acknowledgments}
This study was partially supported by an unrestricted grant from Olympus Co Hamburg, Germany, and by the Forschungszentrum Medizintechnik Hamburg (grant 02fmthh2017), Hamburg, Germany.

RS gratefully acknowledges funding by the Studienstiftung des deutschen Volkes and the G\"unther Elin Krempel foundation. TR receives study support for various  projects from Olympus Co Hamburg, Germany, but declares that there is no conflict to disclose with regards to this project. RW receives funding from Siemens Healthcare, Erlangen, Germany, but declares that there is no conflict to disclose regarding this project.

The authors would like to thank NVIDIA for the donation of a graphics card under the GPU Grant Program. In addition, the authors are grateful toward Hinnerk St\"uben  for his excellent technical support and toward Claus Hilgetag for proofreading and valuable discussions.
\bibliographystyle{model2-names.bst}\biboptions{authoryear}
\bibliography{refs}

\onecolumn

\renewcommand{\thefigure}{S\arabic{figure}}
\setcounter{figure}{0}
\renewcommand{\thesection}{S\arabic{section}}
\setcounter{section}{0}
\renewcommand{\thetable}{S\arabic{table}}
\setcounter{table}{0}
\renewcommand{\thepage}{S-\arabic{page}}
\setcounter{page}{1}
\renewcommand*{\thefootnote}{\fnsymbol{footnote}}
\setcounter{footnote}{0}

\begin{center}
\hspace{1cm}
\textbf{Supplementary Materials to: Multi-scale fully convolutional neural networks for histopathology image segmentation: from nuclear aberrations to the global tissue architecture}\\

\hspace{1cm}

R\"udiger Schmitz\footnote[1]{Corresponding author:
  \texttt{r.schmitz@uke.de}}, Frederic Madesta, Maximilian Nielsen, Jenny Krause, Stefan Steurer, Ren\'e Werner\footnote[7]{Equal contribution.\label{supp:EqualContrib}}, Thomas R\"osch\textsuperscript{\ref{supp:EqualContrib}}
\end{center}

\hspace{1cm}

\begin{multicols}{2}

\section{Base model architecture}
\label{sec:supp_baseline_details}

\begin{figure*}
    \centering
    \includegraphics[width=\textwidth]{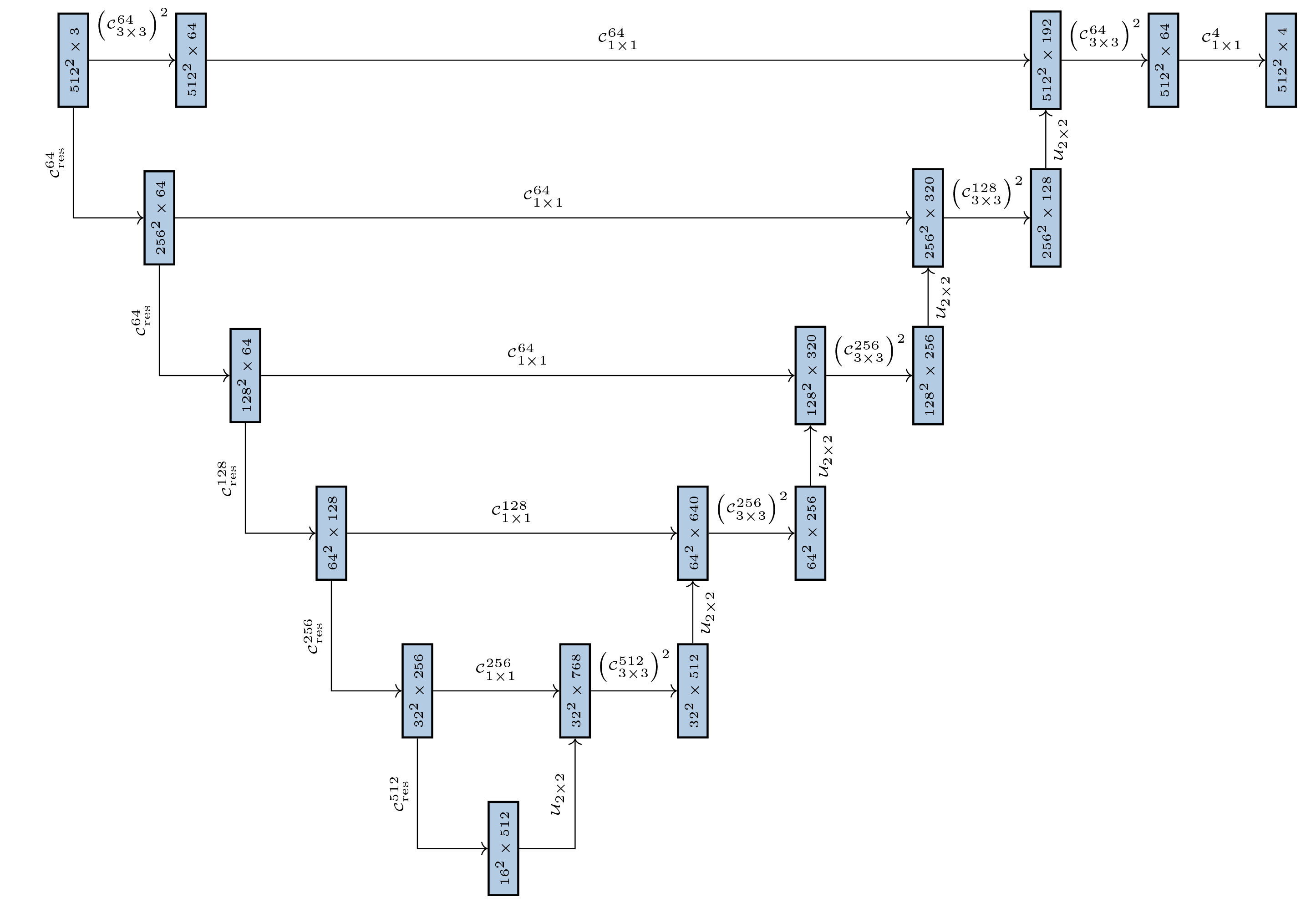}
    \caption{ResNet18-based U-Net architecture (baseline model). For each block the spacial image shape as well as the number of channels are given. Here,
    $\mathcal{C}_{m\times n}^{f_i,f_o}$ denotes a single $m \times n$ convolution with $f_i$ input and $f_o$ output feature maps, followed by ReLU activation.
    $\left(\mathcal{C}_{m\times n}^{f}\right)^l$ shall represent $l$ consecutive $m\times n$ convolutions with $f$ output maps, each followed by a ReLU activation function. For the encoding part, the blocks ($\mathcal{C}_\mathrm{res}^{f}$) of a ResNet18 are used where $f$ denotes number of the respective output feature maps. Each individual block introduces a spatial downscaling by a factor of 2, either through max pooling or strided convolutions. The decoder uses $m\times n$ bilinear upsampling ($\mathcal{U}_{m\times n}$) to enlarge the spacial dimensions.}
    \label{fig:supp_baseline}
\end{figure*}

A detailed view of the baseline model architecture is provided in figure \ref{fig:supp_baseline}.

\section{Establishment of the CAMELYON 2016 macrometastases subset: CAMELYON 2016/MM}
\label{sec:supp_camelyon_subset}

Starting from the original CAMELYON 2016 dataset, we created our subset CAMELYON 2016/MM as follows: First, we sorted all cases by the size of the total tumor region in pixels. Starting from the WSI with the largest total tumor region, we visually checked for the presence of macrometastases, where we additionally excluded cases which contained almost no non-cancerous regions and which had almost no border between healthy and cancerous tissue (as these would not have been well suited for the analysis of segmentation performance). We stopped when $N=20$ samples had been reached. Referring to the original file names in the CAMELYON 2016 dataset, we ended up with the following WSIs in CAMELYON 2016/MM:
``tumor\_009'',
``tumor\_011'',
``tumor\_016'',
``tumor\_026'',
``tumor\_031'',
``tumor\_046'',
``tumor\_047'',
``tumor\_055'',
``tumor\_058'',
``tumor\_068'',
``tumor\_078'',
``tumor\_082'',
``tumor\_085'',
``tumor\_088'',
``tumor\_089'',
``tumor\_090'',
``tumor\_095'',
``tumor\_101'',
``tumor\_102'',
``tumor\_110''.

\section{Baseline model results}
\label{sec:supp_baseline_results}

As described in section \ref{sec:experiments} of the main article, the experiments are repeated three times for the baseline model. The detailed results for each individual repetition are given in table \ref{fig:supp_unet_runs}.

\begin{table*}[]
    \caption{Class-average of the validation Jaccard index for whole and viable tumor at convergence of the validation loss for the baseline model in three independent training runs.}
    \centering
    \begin{tabular}{p{2.0cm} p{1.1cm} p{1.1cm} p{1.1cm} p{1.1cm} p{1.1cm}  }
    \toprule
      &   \multicolumn{5}{c}{\textbf{CV folds}} \\
    \cmidrule(r){2-6}
    \addlinespace
        &   \textbf{0}   &   \textbf{1}   &   \textbf{2}   &   \textbf{3}   &   \textbf{4}    \\
    \hline
    Run 0 &  0.833 & 0.757 & 0.651 & 0.735	& 0.723 \\
     Run 1 &  0.803 & 0.720 & 0.706 & 0.732	& 0.744 \\
      Run 2 &  0.806 & 0.786 & 0.701 & 0.737	& 0.721\\

    \end{tabular}
    \label{fig:supp_unet_runs}
\end{table*}

\section{Statistical significance tests and estimation of the statistical power of the study}
\label{sec:supp_statistical_significance_and_power}

Results from corresponding CV splits are considered as paired measurements.
Use of a paired-sample t-test, however, requires statistical independence of the experiments which is violated in a cross-validation setting.
Therefore, we perform a corrected resampled paired-sample t-test, with the correction as introduced by \citep{NadeauBengio} and discussed by \citep{BouckaertFrank}.
For the reason that all multi-scale architectures can reproduce the behaviour of the baseline U-Net, we use a one-sided resampled paired-sample t-test.
Additionally, in order to correct for multiple testing,
we employ the Benjamini-Hochberg step-up procedure.

Using the results from the baseline model  (cf. section \ref{sec:supp_baseline_results}), we seek to pre-estimate the statistical power of our study: We estimate the standard deviation of the Jaccard index for a single model to be in the range of $0.045$.
Therefore, for the standard deviation of the differences between two models, we assume $\sqrt{2} \cdot 0.045$.
The resampling correction by Nadeau and Bengio can be viewed as an additional factor of $%
\sqrt{\tfrac{1}{n} + \tfrac{n_{val}}{n_{train}}}
/ \sqrt{\tfrac{1}{n}}$ to the standard deviation, where $n$ denotes the number of CV splits and $n_{train}$ and $n_{val}$ are the number of samples in the train and validation set, respectively.
In our experiments, with $n_{train}=40$ and $n_{val}=10$ for PAIP, $n_{train}=8$ and $n_{val}=2$ for BACH and $n_{train}=16$ and $n_{val}=4$ for CAMELYON, this computes to $1.5$. Therefore, we can estimate the power of our study to be equal to the power of a standard paired-sample t-test with a standard deviation of the pairs of $1.5 \cdot \sqrt{2} \cdot 0.045$.
We aim for an improvement over the baseline U-Net of $0.1$ to $0.15$ for the multi-scale models, resulting in a statistical power in between $0.613$ and $0.888$.
For differences between different multi-scale models, which we expect to be around $0.05$, however, our study is clearly underpowered  (statistical power $0.253$).
We therefore conclude that comparisons of multi-scale architectures with respect to the baseline U-Net are amenable to an analysis of statistical significance, whereas differences within the family of multi-scale models are not.

\section{GPU memory footprint analysis}
\label{sec:supp_memory_footprint_results}

For analysis of the GPU memory footprint, we have examined the GPU memory usage of both forward and backward pass as printed by the NVIDIA System Management Interface and performed a linear regression with varying batch size.
Table \ref{tab:supp_regression} provides the detailed results of this.
We interpret the slope of that analysis, i.e. the amount with which the required GPU memory increases when adding another image patch to the batch, as the ``GPU footprint'' of the model and the intercept as its overhead.

\begin{table*}[]
    \caption{Linear regression analysis of the GPU memory usage while training, including forward and backward pass. The memory footprint was read out via the NVIDIA System Management Interface. For the linear regression batches of size 2, 4, 6, 8, 10, 12 and 14 were used.}
        \centering\small
    \begin{threeparttable}
\begin{tabular}{llllrr}
\toprule
\multirow{2}{*}{\textbf{Model}} & \multicolumn{3}{c}{\textbf{Linear regression}}       & \multicolumn{2}{c}{\textbf{Slope}}       \\ \cmidrule(r){2-4} \cmidrule(r){5-6}
                                & \textbf{Slope\tnote{$\dagger$}} & \textbf{Intercept\tnote{$\ddagger$}} & \boldmath{$R^2$} & \textbf{\% U-Net} & \textbf{\% Ens.} \\ \hline
U-Net (scale 1 or 16)                          & 1.066           & 1.512               & 0.994          & 100.0             & 30.4                 \\
U-Net (scale 4) &		1.374	& 1.382	& 0.998 &	128.89	 & 39.19 \\
Ensemble       & 3.506\tnote{\S}           & --                 & --             & 328.9             & 100.0                \\
msY$_{(4)}$-Net                              & 1.422           & 0.914              & 1.000          & 133.4             & 40.6                 \\
msY$_{(16)}$-Net                              & 1.175           & 1.799              & 0.988          & 110.2             & 33.5                 \\
msUI$_{(4)}$-Net                          & 1.389           & 0.538              & 0.997          & 130.3             & 39.6                 \\
msUI$_{(16)}$-Net                              & 1.355           & 0.712              & 0.998          & 127.1             & 38.6                 \\
msY$^2$-Net                    & 1.561           & 0.962              & 1.000          & 146.4             & 44.5                 \\
msY\textit{I}-Net                   & 1.398           & 1.273              & 0.999          & 131.2             & 39.9                 \\
msY$_{(Res34)}^2$-Net     & 1.516           & 1.527              & 0.994          & 142.2             & 43.2                 \\
ms$Y^2$-Net               & 1.436           & 1.669              & 0.984          & 134.7             & 41.0                 \\
msY$_{(16)}$-Net (w/o class. loss)                & 1.175           & 1.799              & 0.988          & 110.2             & 33.5                 \\
Y$_{(16)}$ (w/o class. loss)                 & 1.281           & 1.795              & 0.983          & 120.2             & 36.5                 \\
msY\textit{I}-Net (w/ multiple merges)                        & 1.556           & 1.483              & 0.995          & 146.0             & 44.4                 \\
msY$^2_\text{(Res34)}$-Net (w/ multiple merges)               & 1.608           & 1.532              & 0.995          & 150.8             & 45.9\\
\bottomrule
\end{tabular}
    \begin{tablenotes}
    \item[\S] sum of three individual U-Nets (scales: 1, 4, 16)
    \item[$\dagger$] given in gigabytes per image patch in batch
    \item[$\ddagger$] given in gigabytes
    \end{tablenotes}
    \end{threeparttable}
    \label{tab:supp_regression}
\end{table*}

\section{BACH 2018}

Complementary to table \ref{tab:exp1_bach}, table \ref{tab:supp_exp1_bachmetric} provides the results for the custom metric used in the original BACH 2018 challenge \citep{Aresta2019}. This metric has certain disadvantages, including an unclear mapping of the classes to a rational scale and, importantly, being dominated by contributions of the (trivial) backounground class, and is not used for our study. The results provided here, however, show that the performance of our models lies in the range of the top-performing teams in the original challenge task (1$^{\rm{st}}$ place: 0.69, 2$^{\rm{nd}}$: 0.55, 3$^{\rm{rd}}$: 0.52).

\begin{table*}[]
    \caption{Selected multi-scale architectures versus the baseline U-Net on the BACH 2018 dataset and evaluated by the custom metric used for that challenge.
    	For a quick overview, the best results per split and overall are marked in bold, ignoring differences $<0.005$.
    	}
    \centering\small
    \begin{threeparttable}
    \begin{tabular}{p{1.285cm} p{0.9cm} p{0.9cm} p{0.9cm} P P P P >{\scriptsize}p{0.4cm} p{2.7cm}}
    \toprule
    \textbf{Arch.} & \textbf{Scales} & \textbf{\#\,pms.\tnote{$\dagger$}} & \textbf{Mem.\tnote{$\ddagger$}} & \multicolumn{6}{c}{\textbf{BACH metric}} \\
    \cmidrule(r){5-10}
    \addlinespace
     &   &   &   &   \multicolumn{5}{c}{\textbf{Per CV fold}}  & \textbf{Mean (95\% CI)}  \\
    \hline
\addlinespace[\betweenmodels]
U-Net & 1 & 17.804 & 1.066 & 0.627 & \textbf{0.697} & 0.590 & 0.623 & 0.569
 & 0.621 (0.583, 0.659)
 \\
\addlinespace[\betweenmodels]
Avg. Ens. & 1,\,4,\,16 & 53.412\tnote{\S} & 3.506\tnote{\S} & \textbf{0.641} & 0.692 & 0.688 & 0.801 & 0.682
 & \textbf{0.701 (0.654, 0.748)}
\\
Log. Ens. & 1,\,4,\,16 & 53.412\tnote{\S} & 3.506\tnote{\S} & 0.597 & 0.679 & \textbf{0.703} & 0.761 & 0.555
 & 0.659 (0.594, 0.724)
\\
Maj. Ens. & 1,\,4,\,16 & 53.412\tnote{\S} & 3.506\tnote{\S} & 0.616 & 0.685 & 0.643 & \textbf{0.827} & \textbf{0.692}
 & 0.692 (0.629, 0.756)
\\
\addlinespace[\betweenmodels]
msY\textit{I}-Net
& 1,\,4,\,16 & 40.460 & 1.398 &  0.570 & 0.591 & 0.562 & 0.798 & 0.572
 & 0.619 (0.540, 0.698)
 \\
\addlinespace[\betweenmodels]
msY$^2$-Net & 1,\,4,\,16 & 40.761 & 1.561 & 0.618 & 0.714 & 0.665 & 0.813 & 0.639
 & 0.690 (0.629, 0.751)
\\
    \bottomrule
    \end{tabular}
    \begin{tablenotes}
    \item[\S] sum of three individual U-Nets (scales: 1, 4, 16)
    \item[$\dagger$] in units of one million parameters
    \item[$\ddagger$] GPU memory footprint given in gigabytes per image patch in batch
    \end{tablenotes}
    \end{threeparttable}
    \label{tab:supp_exp1_bachmetric}
\end{table*}

\end{multicols}

\end{document}